\documentclass[journal,onecolumn,12pt]{IEEEtran}
\usepackage{threeparttable}
\usepackage{graphicx}
\usepackage{epstopdf}
\usepackage{amsmath,amssymb}
\usepackage{arydshln}
\usepackage{dsfont}
\usepackage{color}
\usepackage{cases}
\usepackage{bm}
\usepackage{amsfonts}

\usepackage[font=footnotesize]{caption}
\usepackage{indentfirst}
\usepackage{cite}

\usepackage{caption}
\usepackage{algorithm}
\usepackage{algpseudocode}
\usepackage{booktabs}
\usepackage{subfigure}
\usepackage{multirow}
\usepackage{amsthm}
\usepackage{setspace}

  \newcommand{\sys}{}
  \newcommand{\sd}{\mathbf}
\theoremstyle{remark}
\newtheorem{remark}{Remark}
\theoremstyle{}
\newtheorem{theorem}{Theorem}
\theoremstyle{}

\newtheorem{corollary}{Corollary}
\newtheorem{lemma}{Lemma}
\theoremstyle{}
\newtheorem{definition}{Definition}
\theoremstyle{remark}

\theoremstyle{definition}

\makeatletter
\newcommand{\tabcaption}{\def\@captype{table}\caption}

\makeatletter
 \allowdisplaybreaks[4]

\ifCLASSINFOpdf
\else
\fi

\definecolor{newcolor}{rgb}{0.5,0,1}

\newcommand{\rv}[1]{{\color{black}#1}}

\begin{document}

\title{Key Superposition Simultaneously Achieves Security and Privacy in Cache-Aided Linear Function Retrieval}

\author{
Qifa Yan~\IEEEmembership{Member,~IEEE,} and
Daniela Tuninetti~\IEEEmembership{Fellow,~IEEE.}
\thanks{Q. Yan is with the Information Security and National Computing Grid Laboratory, Southwest Jiaotong University, Chengdu 611756, China (e-mail: qifayan@swjtu.edu.cn).
This work was done when Dr. Yan was with the Electrical and Computer Engineering Department of the University of Illinois Chicago, Chicago, IL 60607, USA. }
\thanks{
D. Tuninetti is with the Electrical and Computer Engineering Department of the University of Illinois Chicago, Chicago, IL 60607, USA (e-mail: danielat@uic.edu).}
\thanks{
This paper was presented in part at 2020 IEEE Information Theory Workshop (ITW) \cite{ITW2020} .}
\thanks{
This work was supported in part by NSF Award 1910309.
}
}
\maketitle

\IEEEpeerreviewmaketitle
%

\begin{abstract}

This work investigates the problem of cache-aided content Secure and demand Private Linear Function Retrieval (SP-LFR),
where three constraints are imposed on the system:
(a) each user is interested in retrieving an arbitrary linear combination of the files in the server's library;
(b) the content of the library must be kept secure from a wiretapper who obtains the signal sent by the server; and
(c) \rv{no colluding} subset of users together  obtain information about the demands of the remaining users.
A procedure is proposed to derive an SP-LFR scheme from a given Placement Delivery Array (PDA), which is known to give coded caching schemes with low subpacketization for systems with neither security nor privacy constraints. This procedure uses the superposition of security keys and  privacy keys in both the cache placement and transmitted signal to guarantee content security and demand privacy, respectively. In particular, among all PDA-based SP-LFR schemes, the memory-load pairs achieved by the PDA describing the Maddah-Ali and Niesen's scheme are
Pareto-optimal and have the lowest subpacketization. 
Moreover, the achieved load-memory tradeoff is optimal to within a constant multiplicative gap
except for \rv{the small memory regime (i.e., when the cache size is between $1$ and $2$) and the number of files} is smaller than the number of users.
Remarkably, the memory-load tradeoff does not increase compared to the best known schemes that guarantee \rv{either} only content security in all regimes or only demand privacy in \rv{regime mentioned above}.
\end{abstract}


\section{Introduction}\label{sec:intro}
Coded caching is a known technique to reduce the communication load at peak times in networks with cache-enabled users. A coded caching system is a network composed of a server connected to $K$ users through a shared-link, where the server has access to a library of $N$ files and each user can cache up to $M$ files.  The system operates in two phases. When the network is not congested, the server pushes some contents into the user's cache (placement phase). During peak times,  each user demands one file from the file library, and the server responds by sending a signal over the shared link to enable every user to decode its demand files (delivery phase). Coded caching reduces the worst-case communication load (or just \emph{load} for short in the following) in the delivery phase by designing the cache contents in the placement phase so as to create multicasting opportunities in the delivery phase, regardless of the demanded files.

Coded caching was first proposed by Maddah-Ali and Niesen (MAN)~\cite{Maddah-Ali2014fundamental}. The MAN scheme was proved to achieve the information-theoretical optimal memory-load tradeoff among all uncoded placement schemes when $N\geq K$~\cite{KaiITW}. By removing some redundant transmissions, the optimal load-memory tradeoff among all uncoded placement schemes was completely characterized in~\cite{Yu2019ExactTradeoff,Kai2020Index}.  Improved achievable loads by using coded placement were obtained in~\cite{code:prefetch01,code:prefetch02,code:prefetch03}. Information-theoretic converse bounds were obtained in~\cite{Ramamoorthy2017bound,ImproveBound,QYu2018Factor2}. In particular, the best known converse indicates that the achievable tradeoff in~\cite{Yu2019ExactTradeoff} is to within a constant multiplicative gap of $2.00884$ from the (unknown) information-theoretic optimal tradeoff~\cite{QYu2018Factor2}.
The security of the files in the library against eavesdroppers and the privacy of user demands are both important aspects in practical systems.
In the coded caching literature, information-theoretic security and demand privacy  were considered separately, to the best of our knowledge -- a gap that this works aims to close.

{\bf Content security} was studied in~\cite{secure,Privacy}. In~\cite{secure}, the system needs to protect the content of the library against a wiretapper who obtains the signal sent by the server in the delivery phase. The proposed achievable scheme uses \emph{security keys} shared between the users in order to secure the 
\rv{delivery} signal. 
The security key scheme is optimal to within a multiplicative gap of $17$ when $M\geq\max\big\{1+\frac{(K-N)(N-1)}{KN}, 1\big\}$. \rv{The security guarantee considered in~\cite{Privacy} is stronger than in~\cite{secure}, in the sense that neither the users nor the wiretapper obtain any information that not intended for them.} The proposed scheme creates equal-size shares for each file, such that the cached shares at each user do not reveal any information about the files, and multicast coding is employed on the shares. 
This scheme is optimal to within a multiplicative gap of $16$ when $M\geq\max\big\{1+\frac{N(K-N)}{(N-1)K+N},1\big\}$. 

{\bf Demand privacy} was studied in~\cite{Kai2019Private,Kamath2019,Sneha2019,Aravind2019}, where a user should not gain any information about the index of the file demanded by another user from the 
\rv{delivery} signal. A way to \rv{ensure} demand privacy is to design a delivery scheme to serve virtual users in addition to the actual users~\cite{EliaPrivay}. The observation is that given a known non-private coded caching scheme for $NK$ users (e.g., the schemes in~\cite{Maddah-Ali2014fundamental} or~\cite{Yu2019ExactTradeoff}), the users randomly and privately choose their cache contents from the $NK$ caches without replacement. A given demand vector of the $K$ users is then extended to a demands for $NK$ users (including $K$ real users and $N(K-1)$ virtual users) such that each  file is requested exactly $K$ times. The server sends the signal to satisfy the extended requests of $NK$ users according to the non-private scheme. The demand privacy for the real users is guaranteed since \rv{no} real user can  distinguish the demands \rv{of the real users}  from those of the virtual users. 

In our recent work~\cite{LFR:DPCU}, the demand privacy is enforced against colluding users, that is, \rv{no} subset of users can  obtain information about the demands of the other users, even if they exchange their cache content. This problem was mentioned
in the device-to-device setup~\cite{KaiD2DPrivacy}.  Our proposed scheme in~\cite{LFR:DPCU} uses \emph{privacy keys}, and was inspired by a recent work on cache-aided linear function retrieval in~\cite{Kai2020LinearFunction} that
showed that allowing users to retrieve arbitrary linear combinations of files does not worsen the achievable memory-load tradeoff compared to just retrieving single files. The key idea in~\cite{LFR:DPCU} is that each user caches, in addition to the uncoded cached content as in~\cite{Maddah-Ali2014fundamental},  a privacy key formed as a random linear combination of the content that was not cached in~\cite{Maddah-Ali2014fundamental}. Given the set of files demanded by the users, the server sends multicast signals so that each user can retrieve a specific linear combination of files (related to the stored privacy key). The decoded linear combination together with the privacy key allows each user to \rv{retrieve} the demanded file.
In certain regimes the scheme proposed in~\cite{LFR:DPCU} outperforms the virtual user scheme in~\cite{Kai2019Private} in terms of load-memory tradeoff, and in general has a significant lower subpacketization, \rv{where subpacketization} refers to the minimum file length needed to realize the scheme.

\subsection{Paper Contributions}
In this paper, we investigate content Secure and demand Private Linear Function Retrieval (SP-LFR) systems, where the following constraints are simultaneously imposed:
 \begin{enumerate}
   \item[L:] each user is interested in downloading a \emph{linear} combination of files at the server;
   \item[S:] the files must be kept \emph{secure} from a wiretapper who observes the 
   \rv{delivery} signal from the server\footnote{\rv{This notion of security  is the same as in \cite{secure}, but weaker than that in \cite{Privacy}.}}; and
   \item[P:] \rv{no} subset of users, who may share their cache contents, can  learn information about the 
   \rv{coefficients} of the linear combination demanded by the remaining users thus ensuring \emph{demand} privacy.
 \end{enumerate}
Coded caching systems with only one or two of the above three constraints were investigated \rv{in the literature}:
linear function retrieval~\cite{Kai2020LinearFunction};
content security~\cite{secure};
demand privacy~\cite{Kai2019Private,Aravind2019,Sneha2019,Kamath2019};
linear function retrieval with demand privacy~\cite{LFR:DPCU}.

The main contributions of this paper are as follows.
\begin{enumerate}

  \item We propose to use a \emph{superposition of security keys and privacy keys} to achieve content security and demand privacy simultaneously.  In fact, security keys and privacy keys have been separately used in systems with only security or privacy \rv{constraint}. Each security key was shared by some users~\cite{secure}, while each privacy key was only cached by an individual user~\cite{LFR:DPCU}. In our approach, in the placement phase each user caches the superposition (i.e., sums over the finite field of operation) of security keys  and privacy keys. In the delivery phase, both security and privacy keys are added to the multicast signals. It turns out that this superposition strategy \rv{ensures that}  the content is secure and the demands are kept private. Key superposition neither increases the memory size nor the communication load compared to schemes with either only security or only privacy keys.

  \item We propose a procedure to obtain an SP-LFR scheme from a Placement Delivery Array (PDA) by incorporating the idea of key superpositions and linear function retrieval into the PDA framework originally developed in~\cite{Yan2017PDA}. A PDA characterizes the placement and delivery phases with a single array in coded caching systems with none of the three \rv{constraints} studied in this paper. The advantage of leveraging the PDA framework is that we can conveniently transform all existing PDA structures into SP-LFR schemes. It is well known that, with fixed number of files and memory size, the subpacketization of MAN-based schemes increases exponentially with the number of users, and so do the versions with any one the the three constraints above.  Fortunately, there have been extensive researches on low subpacketization  coded caching schemes in the literature~\cite{FiniteLength2016,Yan2017PDA,YanBipatite2017,ShangguanHypergraph2016,Unified,ChengTcomPDA}, most of which can be characterized by PDA. As a result, characterizing SP-LFR schemes with PDAs makes it possible to conveniently transform those existing low subpacketization structures into SP-LFR schemes.

  \item Among those existing PDAs, the PDA describing the MAN scheme, referred to as MAN-PDA in the following, is of particular \rv{importance}. We show that the memory-load pairs achieved by MAN-PDA in the SP-LFR setup are Pareto-optimal among all PDA-based SP-LFR schemes, and the subpacketization is the lowest subpacketization needed to achieve those points within the PDA framework. No such strong performance gurantee on PDA has been previously reported in the coded caching literature. Remarkably, the MAN-PDA-based SP-LFR scheme does not increase the memory-load tradeoff compared to the best known tradeoff with only security constraint in all regimes, or that with only privacy constraint in some regime. Moreover, the load-memory tradeoff achieved by the MAN-PDA-based SP-LFR scheme is shown to be optimal to within a constant multiplicative gap, except for the regime $M\in[1,2)$ and $K > N$. In addition, the largest constant obtained is 8, which improves the constant $17$ found in~\cite{secure} for systems with only a security constraint.

\end{enumerate}

\subsection{Paper Organization}
Section~\ref{sec:model} introduces the problem formulation.
Section~\ref{sec:PDA:example} reviews the definition of PDA and presents an illustrative example.
Section~\ref{sec:main} summarizes the main results of this paper with detailed proofs deferred to Sections~\ref{sec:scheme}--\ref{sec:gap}.
Section~\ref{sec:numerical} presents some numerical results.
Section~\ref{sec:conclusion} concludes the paper.
Some proofs can be found in Appendix.

\subsection{Notations}
We use $\mathbb{R}^+$ to denote the set of non-negative real numbers and $\mathbb{F}_q$ to denote the finite field with $q$ elements for prime power $q$. For a positive integer $n$, $\mathbb{F}_q^n$ is the $n$ dimensional vector space over $\mathbb{F}_q$, and $[n]$ is the set of the first $n$ positive integers $\{1,2,\ldots,n\}$. \rv{For integers $m,n$ with $m\leq n$, we use $[m:n]$ to denote the set $\{m,m+1,\ldots,n\}$,} and ${n\choose m}$ to denote the binomial coefficient   $\frac{n!}{m!(n-m)!}$, with the convention that ${n\choose m}=0$ if $m > n$. For  a sequence of variables indexed by positive integers $Z_1,Z_2,\ldots,$ and an \rv{integer} index set $\mathcal{S}$, we use the notation $Z_{\mathcal{S}}\triangleq\{Z_{i}:i\in\mathcal{S}\}$.  We use the notations $``+"$ and $``\sum"$ to denote the addition and summations on the real field and on the finite field $\mathbb{F}_q$, respectively, where the meaning of those symbols \rv{is} clear from the \rv{context}. We reserve the notation $``\oplus"$ to denote the Exclusive OR (XOR) operation.
We let $\sd{e}_1,\ldots,\sd{e}_N$ be the standard unit vectors over $\mathbb{F}_q^N$, i.e., $\sd{e}_n$ is the vector  in $\mathbb{F}_q^N$ such that the $n$-th digit is one and all the other digits are zeros. 

\section{System Model}\label{sec:model}
Let $N,K,B$ be positive integers.
An $(N,K)$  caching system consists of a server with $N$ files (\rv{denoted by} $\sys{W}_1,\sys{W}_2,\ldots,\sys{W}_N$) and $K$ users (\rv{denoted by} $1,2,\ldots,K$), where the server is connected to the users via an error-free shared link. The $N$ files are identically and uniformly distributed over $\mathbb{F}_q^B$, for some prime power integer $q$, and for some integer $B$ denoting the file length. 
The system operates in two phases as follows.

\paragraph*{Placement Phase} 
The server privately generates a random variable $\sys{P}$ from some probability space $\mathcal{P}$.
Then it fills the cache of each user $k\in[K]$ with a  \emph{cache function}
$\varphi_k:\mathcal{P}\times\mathbb{F}_q^{NB}\mapsto\mathbb{F}_q^{\rv{\lfloor MB \rfloor}}$.
The cache content of user $k$ is
  \begin{IEEEeqnarray}{c}
   \sys{Z}_k=\varphi_k(\sys{P},\sys{W}_{[N]}),\quad \forall\, k\in[K].\label{eqn:Zk}
   \end{IEEEeqnarray}
The quantity $M$ is the \emph{memory size} at each user.  
\rv{In other words, each user has a memory of size at most $MB$ symbols for some $M\in[0,N]$.}

\rv{We assume that the encoding functions are known to the server and all users, but the randomness $P$ is not available at the users except through the cache content in~\eqref{eqn:Zk}, that is, if any randomness is needed by a user, it must be stored in or computed from the cached content of the user.}

\paragraph*{Delivery Phase} Each user $k\in[K]$ demands $\sd{d}_k=(d_{k,1},\ldots,d_{k,N})^{\top}\in\mathbb{F}_q^N$, which means it aims to retrieve the linear combination
\begin{IEEEeqnarray}{c}
\overline{W}_{\sd{d}_k}\triangleq d_{k,1}\cdot W_1+\ldots+  d_{k,N}\cdot W_N,\label{eqn:Wd}
\end{IEEEeqnarray}
where the addition and multiplication are operated symbol-wise on the finite field $\mathbb{F}_q$.

\rv{We assume that} the file library files $W_{[N]}$, the randomness $P$ and the demands $\sd{d}_1,\ldots,\sd{d}_K$ are independent, that is
\begin{IEEEeqnarray}{c}
H(\sd{d}_{[K]},P,W_{[N]})=\sum_{k=1}^KH(\sd{d}_{k})+H(P)+\sum_{n=1}^NH(W_n),\label{eqn:request}
\end{IEEEeqnarray}
where the base of the logarithm is $q$. 

\rv{The server then} creates a signal $\sys{X}$ by using the \emph{encoding function}
$\phi:  \mathcal{P}\times\mathbb{F}_q^{KN}\times \mathbb{F}_q^{NB}\mapsto\mathbb{F}_q^{\rv{\lfloor RB \rfloor}}$.
The transmitted signal is
\begin{IEEEeqnarray}{rCl}
\sys{X}&=&\phi(\sys{P},\sd{d}_{[K]},\sys{W}_{[N]}).
\end{IEEEeqnarray}
The quantity $R$ is called the \emph{worst-case load} of the system\footnote{\rv{In some of prior works in the literature, $R$ is also referred to as ``rate".  }}.
\rv{In other words, the server sends at most $RB$ symbols for some $R\geq 0$.}

The following conditions must hold for an SP-LFR scheme:
\begin{IEEEeqnarray}{crCl}
\textnormal{[Correctness]}&
H(\overline{W}_{\sd{d}_k}\,|\,\sys{X},\sd{d}_k,Z_k)&=&0,~\forall\,k\in[K],\IEEEeqnarraynumspace\label{eqn:correctness}\\
\textnormal{[Security]}&I(W_{[N]};X)&=&0,\label{eqn:security}\\
\textnormal{[Privacy]}& I(\sd{d}_{{[K]\backslash\mathcal{S}}};\sys{X},\sd{d}_\mathcal{S},Z_\mathcal{S}\,|\,&\sys{W}_{[N]}&)=0,\notag\\
&\forall\,\mathcal{S}&\subseteq&[K],\mathcal{S}\neq \emptyset.\label{eqn:privacy}
\end{IEEEeqnarray}

\paragraph*{Objective}
A memory-load pair $(M,R)\in[1,N]\times \mathbb{R}^+$ is said to be achievable if there exists a scheme such that all the conditions in~\eqref{eqn:correctness}--\eqref{eqn:privacy} are satisfied.
The optimal 
load-memory tradeoff of the system is defined as
\begin{IEEEeqnarray}{c}
R^*(M)=\liminf_{B\rightarrow+\infty}\{R:(M,R)~\textnormal{is achievable}\}.
\end{IEEEeqnarray}

In this paper, our main objective is to characterize the optimal worst-case load-memory tradeoff $R^*(M)$. But we are also interested in the subpacketization level, defined as  the minimum number $B$ needed to realize the scheme.

Throughout the paper, we focus on the case $N\geq 2$,
since for $N=1$ demand privacy is impossible (i.e., there only one possible file to demand).

\begin{remark}[Implications of the Constraints]
\label{remark:privacy}
The security condition in~\eqref{eqn:security} guarantees that a wiretapper, who is not a user in the system and observes the delivery signal,  \rv{obtains no} information about the contents of the files.
The privacy condition in~\eqref{eqn:privacy} guarantees that \rv{no} subset of users who exchange their cache contents  jointly learn any information on the demands of the other users, regardless of the file realizations.
In \cite[Appendix A]{LFR:DPCU}, it was showed that the conditions in~\eqref{eqn:privacy} imply that $I(\sd{d}_{[K]};X\,|\,W_{[N]})=0$. Thus, together with~\eqref{eqn:security}, we obtain
\begin{IEEEeqnarray}{c}
 I(W_{[N]},\sd{d}_{[K]};X)=0,\label{eqn:security:XW}
\end{IEEEeqnarray}
that is, the wiretapper having access to $X$ in fact can not obtain any information on both the contents of the files and the demands of the users. 
In other words, 
the random variable $W_1,\ldots, W_N , \sd{d}_1,\ldots, \sd{d}_{K},\sys{X}$ are mutually independent,
where the crucial resource to ensure independence in~\eqref{eqn:security:XW} is the availability of the randomness $\sys{P}$ at the server. 
Notice that,~\eqref{eqn:security:XW}  implies the security condition~\eqref{eqn:security}, but does not imply the privacy condition~\eqref{eqn:privacy}  due to the presence of $Z_\mathcal{S}$.
\end{remark}

\begin{remark}[Minimum memory size]
It was proved in~\cite{secure} that, in order to \rv{simultaneously} guarantee the correctness condition in~\eqref{eqn:correctness} and the security condition in~\eqref{eqn:security}, the memory size $M$ has to be no less than one. 
\end{remark}

\begin{remark}[Possible notions of privacy]
Different definitions of demand privacy for file retrieval have been used in the literature, such as~\cite{Kai2019Private,Aravind2019,Kamath2019,Sneha2019}. Here we adopt the definition in~\cite{LFR:DPCU}, which is the strongest among the definitions used in the literature and is motivated by the need to \rv{ensure} privacy regardless of the file distribution\footnote{The assumption that files are independent and uniformly distributed is only used in the derivation of converse bounds.}. 
%
\end{remark}

\begin{remark}[Naming convention]
The correctness condition in~\eqref{eqn:correctness} \rv{ensures} that each user can decode its demanded linear function of the files. We shall refer to a scheme that satisfies~\eqref{eqn:correctness} as a Linear Function Retrieval (LFR) scheme. If in addition the scheme satisfies either~\eqref{eqn:security} or~\eqref{eqn:privacy}, we shall refer to it as a Secure LFR (S-LFR) scheme or a Private LFR (P-LFR) scheme, respectively.

If we impose the restriction that the demands $\sd{d}_k\in\{\sd{e}_1,\ldots,\sd{e}_N\}, \ k\in[K]$, then the problem formulation reduces to the case where each user is interested \rv{to retrieve a single} file. Similarly to the linear function retrieval setup, in the \rv{single} file retrieval problem, we refer a scheme that satisfies the correctness condition in~\eqref{eqn:correctness} as a File Retrieval (FR) scheme. If in addition it satisfies~\eqref{eqn:security} or~\eqref{eqn:privacy}, we refer it a Secure FR (S-FR) scheme or a Private FR (P-FR) scheme, respectively. If it satisfies both~\eqref{eqn:security} and~\eqref{eqn:privacy}, we refer to it as SP-FR scheme. In the following, we will refer to the FR scheme in~\cite{Yu2019ExactTradeoff} by YMA scheme, and \rv{to} the LFR scheme in~\cite{Kai2020LinearFunction} as WSJTC scheme.  

Fig.~\ref{fig:venn} shows the relationships between those schemes under those various setups.  Table~\ref{table:compare} summarizes the 
\rv{achievable} corner points of known 
schemes.

  \begin{figure}
  \centering
  \includegraphics[width=0.35\textwidth]{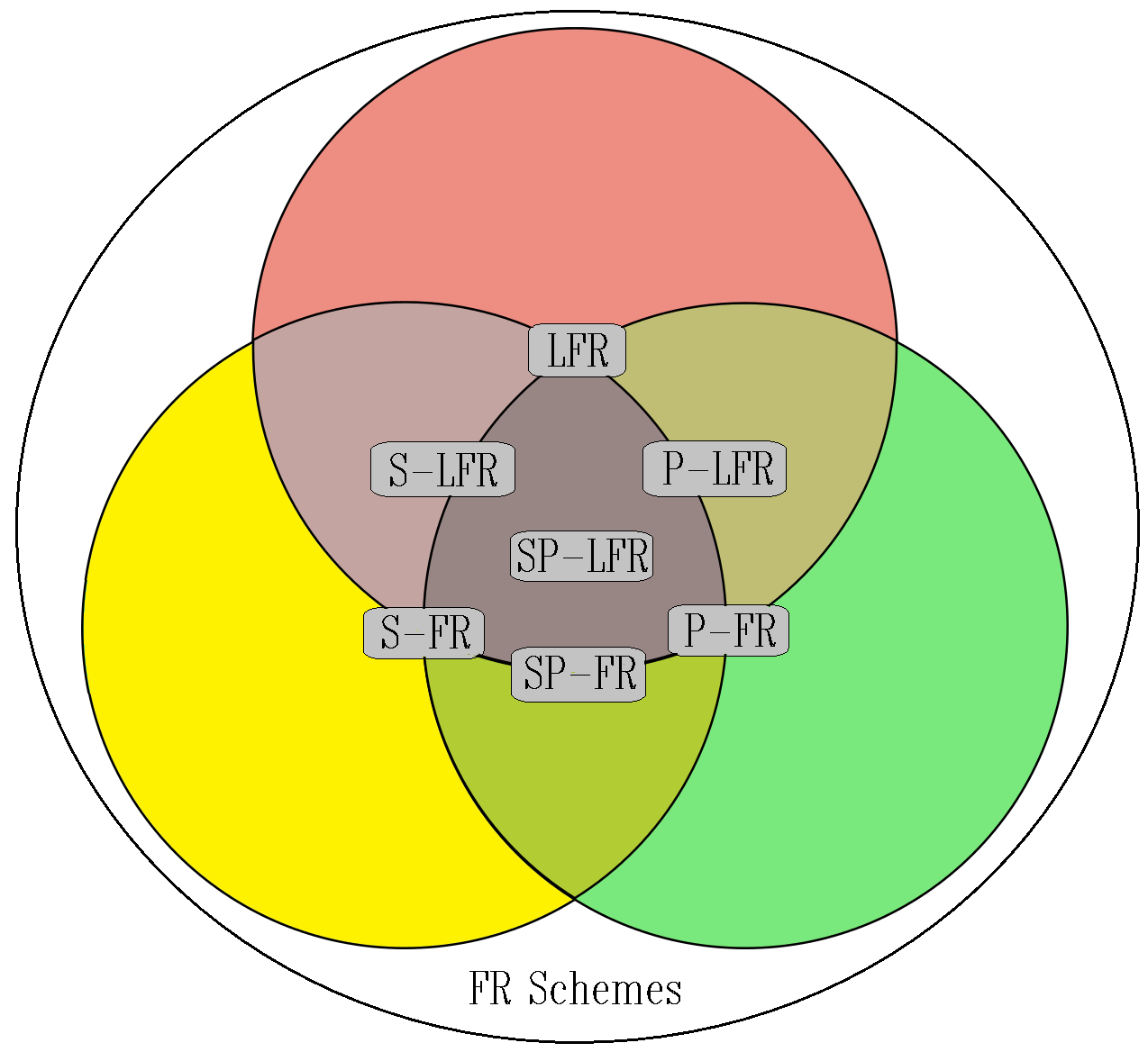}\\
  \caption{Relationships between various caching schemes.}
  \label{fig:venn}
  \end{figure}

\renewcommand\arraystretch{1.8}
\begin{table*}[htbp]\centering
{\begin{threeparttable}
\caption{Performances of known achievable LFR and FR schemes.}
\label{table:compare}
\begin{tabular}{|c|c|c|c|c|c|}
  \hline
  Setup&Scheme&Range of $t$  &Memory $M$  & Load  $R$  & Subpacketization  $B$ \\\hline
    FR & YMA~\cite{Yu2019ExactTradeoff}&$[0:K]$&$\frac{tN}{K}$&$\frac{{K\choose t+1}-{K-\min\{N,K\}\choose t+1}}{{K \choose t}}$&${K\choose t}$\\\hline
   LFR & WSJTC~\cite{Kai2020LinearFunction}&$[0:K]$&$\frac{tN}{K}$&$\frac{{K\choose t+1}-{K-\min\{N,K\}\choose t+1}}{{K \choose t}}$&${K\choose t}$\\\hline
 P-LFR & Privacy Key$^*$\cite{LFR:DPCU}&$[0:K]$&$1+\frac{t(N-1)}{K}$&$\frac{{K\choose t+1}-{K-\min\{N,K\}\choose t+1}}{{K \choose t}}$&${K\choose t}$\\\hline
   \multirow{2}*{P-FR}&Privacy Key$^*$~\cite{LFR:DPCU}&$[0:K]$&$1+\frac{t(N-1)}{K}$&$\frac{{K\choose t+1}-{K-\min\{N-1,K\}\choose t+1}}{{K\choose t}}$ &$K\choose t$\\\cline{2-6}
   &Virtual Users~\cite{Kamath2019}&$[0:KN]$&$\frac{t}{K}$&$\frac{{KN\choose t+1}-{(K-1)N\choose t+1}}{{KN\choose t}}$&${KN\choose t}$\\\hline
   S-FR & Security Key\cite{secure}&$[0:K]$&$1+\frac{t(N-1)}{K}$&$\frac{K-t}{t+1}$&${K\choose t}$\\\hline
\end{tabular}
\begin{tablenotes}
\item[*]
\footnotesize{In the privacy key scheme, the load-memory tradeoff curve  was obtained by taking the lower convex envelope of the points in this row and a trivial point $(M,R)=(0,N)$, which can be achieved with subpacketization $B=1$.}
\end{tablenotes}
\end{threeparttable}}
\end{table*}
\renewcommand\arraystretch{1}

\end{remark}

\section{PDAs and A Toy Example}\label{sec:PDA:example}
Our achievable results are based on the notion of PDA~\cite{Yan2017PDA}, originally introduced  to reduce the subpacketization in the FR setup.
In this section, we first review the definition of PDA, and then give an illustrative example to show our idea to design an SP-FR scheme. How to extend the idea from SP-FR to SP-LFR will be discussed in the rest of the paper.

\subsection{Placement Delivery Array}

\begin{definition}[PDA~\cite{Yan2017PDA}]\label{def:PDA} For given $K,F\in\mathbb{N}^+$ and $Z,S\in\mathbb{N}$,  an $F\times K$ array
  $\mathbf{A}=[a_{i,j}]$, $i\in [F], j\in[K]$, composed of
  $Z$ specific symbols ``$*$"  in each column and some ordinary symbols $1,\ldots, S$,
  each occurring at least once,  is called a $(K,F,Z,S)$ PDA, if, for
  any two distinct entries $a_{i,j}$ and $a_{i',j'}$,   we have
  $a_{i,j}=a_{i',j'}=s$, for some ordinary symbol $s\in[S]$ only if
  \begin{enumerate}
     \item [a)] $i\ne i'$, $j\ne j'$, i.e., they lie in distinct rows and distinct columns; and
     \item [b)] $a_{i,j'}=a_{i',j}=*$, i.e., the corresponding $2\times 2$  sub-array formed by rows $i,i'$ and columns $j,j'$ must be of the following form
  \begin{IEEEeqnarray}{c}
    \left[\begin{array}{cc}
      s & *\\
      * & s
    \end{array}\right]~\textrm{or}~
    \left[\begin{array}{cc}
      * & s\\
      s & *
    \end{array}\right].
  \end{IEEEeqnarray}
   \end{enumerate}
   A PDA is called a $g$-regular PDA if each ordinary symbol occurs exactly $g$ times. \rv{In this case, we call it a $g$-$(K,F,Z,S)$ PDA for short.}
   \end{definition}

\subsection{A Toy Example of SP-FR Scheme from PDA}\label{example:illustrative}

In this subsection, we will derive a SP-FR scheme associated \rv{with} the $2$-$(3,3,1,3)$ PDA 
\begin{IEEEeqnarray}{c}
\mathbf{A}=\left[\begin{array}{ccc}
        * & 1 &2\\
        1 & * &3\\
        2& 3&*
      \end{array}
\right]
\label{eqn:pippo}
\end{IEEEeqnarray}
for an $(N,K)=(4,3)$ caching system. 

Let the four files be $W_1,W_2,W_3,W_4\in\mathbb{F}_2^B$. Firstly, split each file into $F=3$ equal-size packets, i.e., $W_n=(W_{n,1},W_{n,2},W_{n,3})$ for all $n\in[N]$. Each packet is of size $\frac{B}{3}$ bits. The server associates the $i$-th ($i\in[3]$) packet of all files to  the $i$-th row of $\mathbf{A}$, e.g., the packets $\{W_{1,1},W_{2,1},W_{3,1},W_{4,1}\}$ are associated \rv{with} the first row. Users $1$, $2$ and $3$ are associated \rv{with} columns $1$, $2$ and $3$ of $\mathbf{A}$, respectively.

In the placement phase, the server first generates $S=3$ \emph{security keys}, and $K(F-Z)=6$ \emph{privacy keys} as follows.
   \begin{itemize}

     \item The security keys, denoted by $V_1,V_2$ and $V_3$,  are associated \rv{with} the ordinary symbols $1$, $2$ and $3$, respectively. They are independent and uniformly chosen from $\mathbb{F}_2^{B/3}$.

     \item The privacy keys, denoted by $\{T_{i,j}:a_{i,j}\neq *\}$ where $T_{i,j}$ \rv{belongs to $\mathbb{F}_2^{B/3}$} \rv{and} is associated \rv{with} the entry $a_{i,j}$, are generated as follows. The server first generates $K=3$ random vectors $\sd p_1,\sd p_2,\sd p_3$, one for each user (or column of $\mathbf{A}$),  which are independently and uniformly chosen from the vectors of \rv{$\mathbb{F}_2^N$ (in this example, $N=4$)}, i.e.,
         \begin{IEEEeqnarray}{rCl}
         \sd p_{j}=(p_{j,1},p_{j,2},p_{j,3},p_{j,4})^{\top}&\sim&\textnormal{Unif}\big\{\mathbb{F}_2^4\big\},~\forall\,j\in[3].\IEEEeqnarraynumspace
         \end{IEEEeqnarray}
     Then the $6$ keys, denoted as $T_{2,1},$ $T_{3,1},$ $T_{1,2},$ $T_{3,2},$ $T_{1,3},$ $T_{2,3}$, are given as
     \begin{IEEEeqnarray}{rCl}
     T_{i,j}&=&p_{j,1}W_{1,i}\oplus p_{j,2}W_{2,i}\oplus p_{j,3}W_{3,i}\oplus p_{j,4}W_{4,i},\notag\\\quad &&\quad\quad\forall\,(i,j)\in[3]\times[3]~\textnormal{s.t.}~a_{i,j}\neq *.\label{eqn:exam:T}
     \end{IEEEeqnarray}

   \end{itemize}
The server \rv{populates} the cache of users $1$, $2$, $3$ according to $\mathbf{A}$ as follows.  For each $i\in[3]$, if $a_{i,j}=*$, user $j$ stores $\{W_{1,i},W_{2,i},W_{3,i},W_{4,i}\}$ in its cache, or else, user $j$ stores the superposition key $T_{i,j}\oplus V_{a_{i,j}}$. In this example, the contents $Z_1$, $Z_2$ and $Z_3$ are listed in Table~\ref{table:Z}. \rv{Notice that, the variables $V_1,V_2,V_3,\sd{p}_1,\sd{p}_2,\sd{p}_3$ and $T_{i,j}$ in~\eqref{eqn:exam:T} are not cached by the users, and hence they are not known by the users.}

\begin{table}[htbp]\centering
\caption{The cache contents created according to $\mathbf{A}$ in~\eqref{eqn:pippo}.}\label{table:Z}
\begin{tabular}{ccc}
  \toprule
 $Z_1$&$Z_2$&$Z_3$\\\hline
 $W_{[4],1}$&$T_{1,2}\oplus V_1$&$T_{1,3}\oplus V_2$\\
 $T_{2,1}\oplus V_1$&$W_{[4],2}$& $T_{2,3}\oplus V_3$\\
 $T_{3,1}\oplus V_2$&$T_{3,2}\oplus V_3$&$W_{[4],3}$\\
 \bottomrule
\end{tabular}
\end{table}

In the delivery phase, assume that  \rv{user} 
$k$ demands file $W_k, k\in[3]$, i.e., \rv{the} demands are given by
 \begin{IEEEeqnarray}{c}
 \sd{d}_1=(1,0,0,0)^\top,\,
 \sd{d}_2=(0,1,0,0)^\top,\,
 \sd{d}_3=(0,0,1,0)^\top.\IEEEeqnarraynumspace
 \end{IEEEeqnarray}
The server then creates three vectors $\sd{q}_1,\sd{q}_2,\sd{q}_3$ as
\begin{IEEEeqnarray}{rCl}
[\sd{q}_1,\sd{q}_2,\sd{q}_3]&=&[\sd p_1\oplus \sd d_1,\sd p_2\oplus \sd d_2,\sd p_3\oplus \sd{d}_3]\\
&=&\left[\begin{array}{ccc}
          p_{1,1}\oplus 1 & p_{2,1} & p_{3,1}  \\
          p_{1,2} & p_{2,2}\oplus 1 & p_{3,2} \\
          p_{1,3} & p_{2,3}  & p_{3,3}\oplus 1 \\
          p_{1,4} & p_{2,4} & p_{3,4}
        \end{array}
\right].\label{eqn:exam:Q}
\end{IEEEeqnarray}
Then, three signals $Y_1,Y_2$ and $Y_3$, associated \rv{with} the ordinary symbols $1,2,3$ in $\mathbf{A}$ in~\eqref{eqn:pippo} are  created  as
\label{eqn:exam:Y}
\begin{IEEEeqnarray}{rCl}
Y_1=V_1\oplus (W_{1,2}\oplus T_{2,1})\oplus (W_{2,1}\oplus T_{1,2}),\\
Y_2=V_2\oplus (W_{1,3}\oplus T_{3,1})\oplus (W_{3,1}\oplus T_{1,3}),\\
Y_3=V_3\oplus (W_{2,3}\oplus T_{3,2})\oplus (W_{3,2}\oplus T_{2,3}).
\end{IEEEeqnarray}
The server sends the signal
\begin{IEEEeqnarray}{C}
X \triangleq (\sd{q}_1,\sd{q}_2,\sd{q}_3,Y_1,Y_2,Y_3)
\end{IEEEeqnarray}
to the users.
Notice that, by~\eqref{eqn:exam:T}, for all $i,j\in[3]$,
\begin{IEEEeqnarray}{c}
W_{j,i}\oplus T_{i,j}=(p_{j,j}\oplus 1)W_{j,i}\oplus\bigoplus_{n\in[4]\backslash \{j\}}p_{j,n}W_{n,i}.\label{eqn:exam:privacy:packets}
\end{IEEEeqnarray}
Thus, by~\eqref{eqn:exam:Q}, the coefficients of the packets involved in~\eqref{eqn:exam:privacy:packets} are given by $\sd{q}_j$ and are known by the users.

The users decode their demanded packets as follows. User $k\in[3]$ has $W_{1,k},W_{2,k},W_{3,k},W_{4,k}$ in its cache from Table~\ref{table:Z}. Thus, it can compute $W_{j,k}\oplus T_{k,j}$ for any $j\in[3]\backslash\{k\}$.  Moreover, notice from  Table~\ref{table:Z}, user $1$ has the keys  $V_1\oplus T_{2,1}$ and $V_2\oplus T_{3,1}$ in its cache, so user $1$ can decode $W_{1,2}$ and $W_{1,3}$ from $Y_1$ and $Y_2$, respectively. Similarly,   user $2$ can decode $W_{2,1}$ and $W_{2,3}$ from $Y_1$ and $Y_3$, respectively; and user $3$ can decode $W_{3,1}$ and $W_{3,2}$ from $Y_2$ and $Y_3$, respectively.

The privacy is guaranteed since from the users' viewpoint, the vectors $\sd{q}_1,\sd{q}_2,\sd{q}_3$ are 
independently and uniformly distributed over $\mathbb{F}_2^4$.
The security is guaranteed since each coded packet is accompanied by a unique security key.

Finally, each  packet is of size $\frac{B}{3}$ bits. Each user caches $6$ packets, and the server sends $3$ packets. The vectors $\sd{q}_1,\sd{q}_2,\sd{q}_3$  can be sent in $H(\sd{q}_1,\sd{q}_2,\sd{q}_3)=3\times 8=24$ bits, which does not scale with the file size $B$. Therefore, the scheme achieves $(M,R)=(2,1)$.


\section{Main Results}\label{sec:main}
In this section we present our main results, which is a generalization of the example in Section \ref{example:illustrative}. Details and proofs are deferred to Sections~\ref{sec:scheme}--\ref{sec:gap}.

\subsection{PDA Based SP-LFR Schemes}
With any fixed PDA, we will construct an associated SP-LFR scheme.
The following theorem summarizes the performance of PDA-based SP-LFR scheme, which will be proved by presenting and analyzing the construction in Section~\ref{sec:scheme}.
\begin{theorem}\label{thm:PDA} For any $(N,K)$ system and a given $(K,F,Z,S)$ PDA $\mathbf{A}$, there exists an associated SP-LFR scheme that achieves the  memory-load pair
\begin{IEEEeqnarray}{c}
\big(M_{\mathbf{A}},R_{\mathbf{A}}\big)=\bigg(1+\frac{Z(N-1)}{F}\,,\,\frac{S}{F}\bigg).\label{eqn:PDA:MR}
\end{IEEEeqnarray}
with subpacketization $F$.
\end{theorem}

By the procedure described in Section~\ref{sec:scheme}, we can easily obtain SP-LFR schemes from existing PDA constructions. In particular, the following PDA from the Maddah-Ali and Niesen scheme is important, referred to as MAN-PDA.

\begin{definition}[MAN-PDA]\label{def:MNPDA}
For any integer $t\in[0:K]$, define the set $\mathbf{\Omega}_t\triangleq\{\mathcal{T}\subseteq[K]:|\mathcal{T}|=f\}$; \rv{sometimes we shall index the elements of $\mathbf{\Omega}_t$ as  $\mathbf{\Omega}_t=\{\mathcal{T}_i\}_{i=1}^{ K \choose t}$}.
Also, choose an arbitrary bijective function $\kappa_{t+1}$ from $\mathbf{\Omega}_{t+1}$
to the set $\big[{K \choose {t+1}}\big]$.
Then, define the
	array $\mathbf{A}_t=[a_{i,j}]$ as
	\begin{IEEEeqnarray}{c}
		a_{i,j}\triangleq \left\{\begin{array}{ll}
			*, &\textnormal{if}~j\in\mathcal{T}_i \\
			\kappa_{t+1}(\{j\} \cup \mathcal{T}_{i}), &\textnormal{if}~j\notin\mathcal{T}_i
		\end{array}
		\right..
	\end{IEEEeqnarray}
\end{definition}

\begin{corollary}\label{cor:MAN}
Let $R_{\textnormal{MAN}}(M)$ be the 
lower convex envelope of the following points
\begin{IEEEeqnarray}{rCl}
(M_t, R_t)= \left(1+\frac{t(N-1)}{K}, \frac{K-t}{t+1}\right),\quad t\in[0:K],\IEEEeqnarraynumspace\label{eqn:MRt}
\end{IEEEeqnarray}
then $R_{\textnormal{MAN}}(M)$ is achievable in an $(N,K)$ SP-LFR system, where the point $(M_t,R_t)$ can be achieved with subpacketization  ${K\choose t}$. 
\end{corollary}

\begin{IEEEproof}
It was proved in~\cite{Yan2017PDA} that $\mathbf{A}_t$ in Definition \ref{def:MNPDA} is a $(K,{K\choose t},{K-1\choose t-1},{K\choose t+1})$ PDA. Thus, the achievability of the point $(M_t,R_t)$ directly follows from Theorem~\ref{thm:PDA}. Moreover, the lower convex envelope of the points in~\eqref{eqn:MRt} can be achieved by memory-sharing technique~\cite{Maddah-Ali2014fundamental}.
\end{IEEEproof}

\begin{remark}[Known setups subsumed by our construction]
The PDA $\mathbf{A}$ in Subsection~\ref{example:illustrative} is an MAN-PDA. 
In general, for the MAN-PDA-based SP-LFR scheme presented in Section~\ref{sec:scheme}, we have the following.
  If the security keys are removed (by setting  $V_1,\ldots,V_S$ to be  zero vectors),  then the scheme \rv{reduces} to the privacy key scheme in~\cite{LFR:DPCU}\footnote{In the case $N\leq K$ and $t\leq K-N$, some redundant signals are removed in the privacy key scheme \cite{LFR:DPCU} in the P-FR setup. Those signals can not be removed in SP-LFR or S-FR setups due to the use of security keys.}.
  If the privacy keys 
   are removed (by setting $\sd{p}_1,\ldots,\sd{p}_K$ to be zero vectors), and  users demands a \rv{single} file,  the scheme \rv{reduces} to the security key scheme in~\cite{secure}.
  If both the security  and privacy keys are removed, and users demands a \rv{single} file,  then the scheme \rv{reduces} to the MAN scheme~\cite{Maddah-Ali2014fundamental}.
Notice that from Table~\ref{table:compare}, among the best known memory tradeoffs in various previous setups, the worst tradeoff is \rv{for} S-FR, which \rv{coincides with} $R_{\textnormal{MAN}}(M)$. Thus, due to the superposition of security and privacy keys, neither the memory size nor the communication load is increased in the more restrictive SP-LFR setup.  More generally, the following facts hold for \rv{a $(K,F,Z,S)$} PDA-based SP-LFR schemes:
\begin{itemize}
  \item If the security keys  are removed, the SP-LFR scheme \rv{reduces} to a P-LFR scheme, \rv{achieving the same load-memory tradeoff in~\eqref{eqn:PDA:MR}};
  \item If the privacy keys are removed, the SP-LFR scheme \rv{reduces} to an S-LFR schemes, \rv{achieving the same load-memory tradeoff in~\eqref{eqn:PDA:MR}}; and
  \item If both the security  and privacy keys are removed, the SP-LFR scheme \rv{reduces} to  an LFR scheme, \rv{achieving the load-memory pair $(M,R)=(\frac{NZ}{F},\frac{S}{F})$.} In  the FR setup, it \rv{reduces} to the PDA-based FR scheme  in~\cite{Yan2017PDA}.
\end{itemize}
\rv{In addition, when users demand a single  file, the scheme reduces to an SP-FR scheme, achieving the same load-memory tradeoff as in~\eqref{eqn:PDA:MR}. In particular, conclusions similar to those in Theorems \ref{thm:MAN:opt:PDA} and \ref{thm:gap} on the optimality of the MAN-PDA-based scheme hold for the SP-FR setup, as the derivation of these results did not involve LFR demands.}
\end{remark}

\begin{remark}[Subpacketization]
In the P-FR setup, by using the idea of virtual users, it was showed in~\cite{Aravind2019} that one can construct a coded caching scheme for an $(N,K)$ system from any given $(NK,F,Z,S)$ PDA, achieving the memory-load pair $\big(\frac{NZ}{F},\frac{S}{F}\big)$ with subpacketization $F$. In our approach, for a system with $K$ users, we only need a PDA with $K$ columns. \rv{In many PDA constructions,  for example, in the MAN-PDAs and  the ones in~\cite{Yan2017PDA}, the subpacketization parameter $F$ increases exponentially in $K$ (see \cite{Yan2017PDA} for details). }
Thus, in addition to achieve security, our construction has the advantage of avoiding the boost in subpacketization compared to~\cite{Aravind2019}. 
\end{remark}

The following two subsections \rv{show} the optimality of the MAN-PDA-based SP-LFR scheme in two senses.  Theorem~\ref{thm:MAN:opt:PDA} shows the optimality of memory-load pairs and subpacketizations among all PDA-based SP-LFR schemes.   Theorem~\ref{thm:gap} compares the load-memory tradeoff  with the optimal load-memory tradeoff in information-theoretical sense.

\subsection{Lower Bound \rv{for} PDA-based SP-LFR Schemes and Optimality of the MAN-PDA}
\begin{theorem}\label{thm:MAN:opt:PDA} Given a $(K,F,Z,S)$ PDA, 
if the associated SP-LFR scheme achieves a memory-load pair $(M,R)$,
then necessarily
\begin{IEEEeqnarray}{c}
R\geq \frac{K(N-M)}{N-1+K(M-1)}=\left.\frac{K-x}{x+1}\right|_{x = K\frac{M-1}{N-1}}.\label{eqn:PDA:bound}
\end{IEEEeqnarray}
In particular, the memory-load pairs $\{(M_t,R_t):t\in[0:K]\}$ \rv{in Corollary~\ref{cor:MAN}}
satisfy ~\eqref{eqn:PDA:bound} with  equality. Moreover, if $M=M_t$ and $R=R_t$ for some $t\in[0:K]$, then $F\geq{K\choose t}$.
\end{theorem}

Theorem~\ref{thm:MAN:opt:PDA} shows that the  memory-load pairs $\{(M_t,R_t):t\in[0:K]\}$ in~\eqref{eqn:MRt} are in fact Pareto-optimal among all PDA-based SP-LFR schemes, and that the MAN-PDA achieves these points with the lowest subpacketization. Thus, it is impossible \rv{for a PDA-based scheme} to decrease the subpacketization without any \rv{increase} in memory or load.
 \rv{There} have been various results in PDA or equivalent forms in coded caching literature, which pursue low subpacketizations~\cite{FiniteLength2016,Yan2017PDA,YanBipatite2017,ShangguanHypergraph2016,Unified}. From Theorem~\ref{thm:PDA}, we can \rv{analyze} the performance of those schemes when they are used to derive SP-LFR schemes by following the procedure described in Section~\ref{sec:scheme}. As an example, we evaluate the performance of the constructions in~\cite{Yan2017PDA} in Corollary \ref{corollary:PDA}.

Notice that in the FR setup, \rv{the} MAN-PDA-based FR scheme was showed to be optimal among all regular PDA-based FR schemes, and it has the optimal subpacketization among such FR schemes  achieving the same memory-load pairs~\cite{Yan2017PDA}. In our SP-LFR setup, the optimality of MAN-PDA-based SP-LFR scheme in Theorem \ref{thm:MAN:opt:PDA} is stronger than its counterpart in~\cite{Yan2017PDA} in the sense that the optimality is among all PDA-based SP-LFR schemes. In the other setups imposing \rv{the correctness condition \eqref{eqn:correctness}  and one of the two conditions (i.e., security \eqref{eqn:security} and privacy \eqref{eqn:privacy}),} no such results on PDA have been reported, to the best of our knowledge.

\begin{corollary}\label{corollary:PDA} For any $(N,K)$ caching system with $K > 2$, for any $t\in[2:K-1]$ such that $t\,|\,K$ or $(K-t)\,|\,K$, there exists a SP-LFR scheme  achieving the memory-load pair  $(M,R)=\big(1+\frac{t(N-1)}{K},\frac{K-t}{t}\big)$ with subpacketization $\frac{t}{K}\big(\frac{K}{\min\{t,K-t\}}\big)^{\min\{t,K-t\}}$.
\end{corollary}

\begin{IEEEproof} By the PDA construction in~\cite{Yan2017PDA}, for any $m,r\in\mathbb{N}^+$ such that, $r\geq 2$, there exists
\begin{enumerate}
  \item [P$1$:] a $(r(m+1),r^m,r^{m-1},r^{m+1}-r^m)$ PDA;
  \item [P$2$:] a $(r(m+1),(r-1)r^m,(r-1)^2r^{m-1},r^m)$ PDA.
\end{enumerate}
Therefore, we have
\begin{enumerate}
  \item if $t\,|\,K$, let $r= \frac{K}{t}$, $m= t-1$, then by Theorem~\ref{thm:PDA}, the associated SP-LFR scheme in P$1$ achieves the memory-load pair $(M,R)=\big(1+\frac{N-1}{r},r-1\big)$ with subpacketization $r^m$;
  \item if $(K-t)\,|\,K$, let $r= \frac{K}{K-t}$, $m=K- t-1$, then by Theorem~\ref{thm:PDA}, the associated SP-LFR scheme in P$2$ achieves the memory-load pair $(M,R)=\big(1+\frac{(r-1)(N-1)}{r},\frac{1}{r-1}\big)$ with subpacketization $(r-1)r^m$.
\end{enumerate}
In both cases, by plugging $r$ and $m$ into the expressions of $M,R,F$
we conclude that the scheme achieves the memory-load pair $(M,R)=\big(1+\frac{t(N-1)}{K},\frac{K-t}{t}\big)$ with subpacketization $\frac{t}{K}(\frac{K}{t})^{t}$ (if $t\,|\,K$) or $\frac{t}{K}(\frac{K}{K-t})^{K-t}$ (if $(K-t)|K$).
\rv{Finally, the equality} 
\begin{IEEEeqnarray}{c}
\min\{t,K-t\}=\left\{\begin{array}{ll}
                       t, &\textnormal{if}~t\,|\,K \\
                       K-t, & \textnormal{if}~(K-t)\,|\,K
                     \end{array}
\right.
\end{IEEEeqnarray}
\rv{concludes the proof.}
\end{IEEEproof}

\subsection{Gap Results \rv{for the} MAN-PDA-based SP-LFR Scheme}

\begin{theorem}\label{thm:gap}For an $(N,K)$ caching system, the ratio of the achieved load of the MAN-PDA-based SP-LFR scheme $R_{\textnormal{MAN}}(M)$ and the optimal load $R^*(M)$ satisfies
\begin{enumerate}
  \item \rv{if} $N\geq K$ and $M\in[1,N)$,
  \begin{IEEEeqnarray}{c}
  \frac{R_{\textnormal{MAN}}(M)}{R^*(M)}\leq \left\{\begin{array}{ll}
1,&\textnormal{if}~K=1\\
                                                    2, &\textnormal{if}~N=K=2  \\
                                                    6.02652, &\textnormal{if}~N=K\geq 3  \\
                                                    5.0221, & \textnormal{if}~N=K+1 \\
                                                    4.01768, &\textnormal{if}~N\geq K+2
                                                  \end{array}
  \right.;\label{eqn:gap1}
  \end{IEEEeqnarray}
  \item \rv{if} $N < K$ and $M\in[2,N)$,
  \begin{IEEEeqnarray}{c}
  \frac{R_{\textnormal{MAN}}(M)}{R^*(M)} < 8.\label{eqn:gap2}
  \end{IEEEeqnarray}
\end{enumerate}
\end{theorem}

\begin{remark}[Improvement over S-FR]
%
%
It was proved in~\cite{secure} for an S-FR system that $R_{\textnormal{MAN}}(M)$ is to within a constant multiplicative gap of $17$ of the optimal load in the regime $\max\big\{1+\frac{(K-N)(N-1)}{KN},1\big\}\leq M\leq N$. Theorem~\ref{thm:gap} improves the constant over all interval $[1,N]$ to the gaps in~\eqref{eqn:gap1} for $N\geq K$ and to $8$ over $[2,N]$ for $N < K$  in~\eqref{eqn:gap2}.

It is \rv{worth} pointing out that the converse we use in Section~\ref{sec:gap} to derive the gap uses neither the privacy condition~\eqref{eqn:privacy} nor the linear function retrieval condition. This means that  Theorem~\ref{thm:gap} also improves the gap of the S-FR system~\cite{secure}. Since the \rv{converse} bound for an S-FR system also works in an SP-LFR system, the gap result in~\cite{secure} also indicates that in the regime $N < K$ and $1+\frac{(K-N)(N-1)}{NK}\leq M\leq 2$, the load $R_{\textnormal{MAN}}(M)$ is optimal to within the constant gap of $17$.

At present, the gap remains unbounded (i.e., it scales with $N$) only for $K > N$ and $1\leq M\leq1+\frac{(K-N)(N-1)}{KN}$. From Table \ref{table:compare} and Corollary \ref{cor:MAN}, at the point $M=1$ (by setting the parameter $t=0$ in the privacy/security key schemes \rv{in}~\eqref{eqn:MRt}),  it can be observed that in the P-FR (resp. P-LFR) setup, the load $R=\min\{N-1,K\}$ (resp. $R=\min\{N,K\}$) is achievable, while in the S-FR setup  and in our SP-LFR setup the best known achievable load is $R=K$. Thus, it seems that the larger load  when $K > N$ is mainly caused by the security condition; closing the gap in small memory regime \rv{$M\in\big[1,1+\frac{(K-N)(N-1)}{KN}\big]$} when $K > N$ is an open problem in the S-FR setup~\cite{secure}.
\end{remark}

\section{PDA Based SP-LFR Schemes (Proof of Theorem~\ref{thm:PDA})}\label{sec:scheme}
In this section, we first present a SP-LFR scheme for any given PDA, and then prove Theorem~\ref{thm:PDA} by verifying its correctness, security, privacy and performance.

Given a $(K,F,Z,S)$ PDA $\mathbf{A}=[a_{i,j}]_{F\times K}$, the server partitions each file $\sys{W}_n$ into $F$ equal-size packets
\begin{IEEEeqnarray}{c}
\sys{W}_{n}=(\sys{W}_{n,1},\sys{W}_{n,2},\ldots,\sys{W}_{n,F}),\quad\forall~n\in[N].\label{eqn:pda:packets}
\end{IEEEeqnarray}

The server associates the $i$-th packet of the files with the $i$-th row, and user $j$ with the $j$-th column of $\mathbf{A}$.
Then the system operates as follows.

\paragraph*{Placement Phase} The server first generates $S$ \emph{security keys} and $K(F-Z)$ \emph{privacy keys} as follows.
\begin{itemize}
  \item The $S$ security keys, denoted by $V_1,V_2,\ldots,V_S$, are associated \rv{with} the ordinary symbols $s=1,2,\ldots,S$ respectively. They are independently and uniformly chosen from $\mathbb{F}_q^{B/F}$;
  \item The privacy keys are generated as follows.
  The server first  generates $K$ i.i.d. random vectors 
   from  $\mathbb{F}_q^N$, namely
\begin{IEEEeqnarray}{c}
\sd{p}_j\triangleq(p_{j,1},\ldots,p_{j,N})^\top\sim\textnormal{Unif}\big\{\mathbb{F}_q^N\big\},\quad \forall\,j\in[K].\IEEEeqnarraynumspace\label{u:dist}
\end{IEEEeqnarray}
Then the $K(F-Z)$ privacy keys, denoted by $\{T_{i,j}: (i,j)\in[F]\times [K], a_{i,j}\neq *\}$, are generated as follows
\begin{IEEEeqnarray}{rCl}
T_{i,j}\triangleq \mathop\sum\limits_{n\in[N]}p_{j,n}\cdot W_{n,i},\label{eqn:privacy:key}
\end{IEEEeqnarray}
for all $(i,j)\in[F]\times[K]$ such that $a_{i,j}\neq *$, where the key $T_{i,j}$ is associated \rv{with} the entry $a_{i,j}$.
\end{itemize}

The server then fills the cache of user $k\in[K]$ as
\begin{subequations}
\label{eqn:cache}
\begin{IEEEeqnarray}{rCl}
\sys{Z}_k&=&\Big\{\sys{W}_{n,i}: i\in[F],a_{i,k}=*,n\in[N]\Big\}\label{eqn:cache:a}\\
&&\bigcup\Big\{V_{a_{i,k}}+ T_{i,k}: i\in[F],a_{i,k}\neq *\Big\}.\label{eqn:cache:b}
\end{IEEEeqnarray}
\end{subequations}
The random variable $\sys{P}$  is given  by
\begin{IEEEeqnarray}{c}
P=\big(V_{[S]},\sd{p}_{[K]}\big).
\end{IEEEeqnarray}

\paragraph*{Delivery Phase}
After receiving the user demands $\sd{d}_{[K]}$, the server generates $K$ vectors
\begin{IEEEeqnarray}{c}
\sd{q}_k=\sd{p}_k+ \sd{d}_{k}=(q_{k,1},q_{k,2},\ldots,q_{k,N})^{\top},~ \forall\, k\in[K].\IEEEeqnarraynumspace \label{eqn:bk}
\end{IEEEeqnarray}
 The server generates $S$ signals
\begin{IEEEeqnarray}{c}
\sys{Y}_{s}\triangleq V_s
+ \mathop{\sum}\limits_{\substack{(i,j)\in[F]\times[K]\\a_{i,j}=s}} \
  \mathop{\sum}\limits_{n\in[N]} \sys{q}_{j,n}\cdot\sys{W}_{n,i},\quad \forall\, s\in[S]. \IEEEeqnarraynumspace \label{eqn:pda:Ys}
\end{IEEEeqnarray}
Then the server
sends
\begin{IEEEeqnarray}{c}
\sys{X}=\big(\sd{q}_{[K]},Y_{[S]}\big).
\label{eqn:X}
\end{IEEEeqnarray}


\paragraph*{Correctness}
Let
\begin{IEEEeqnarray}{c}
\overline{W}_{\sd{d}_k,i}\triangleq \sum_{n\in[N]} d_{k,n}\cdot W_{n,i},\quad \forall\, k\in[K], i\in[F].\label{eqn:Wdi}
\end{IEEEeqnarray}
We need to show that user $k\in[K]$ can obtain all the packets $\{\overline{W}_{\sd{d}_k,i}:i\in[F]\}$ by~\eqref{eqn:Wd}.
By~\eqref{eqn:cache:a}, for $h\in[F]$ such that $a_{h,k}=*$, user $k$ can compute $\overline{W}_{\sd{d}_k,h}$ directly from the packets in its cache. Thus,  user $k\in[K]$ only needs to decode $\overline{W}_{\sd{d}_k,h}$ such that $a_{h,k}\neq*$.

By the definition of PDA, $a_{h,k}\in[S]$. \rv{For} $s\triangleq a_{h,k}$, we prove that the packet $\overline{W}_{\sd{d}_k,h}$ can be decoded from the signal $Y_s,\sd{q}_{[K]}$ and the cache contents in $Z_k$. In fact, by~\eqref{eqn:bk},~\eqref{eqn:pda:Ys} and~\eqref{eqn:Wdi}, $Y_s$ can be written as
\begin{subequations}\label{decode:Ys}
\begin{IEEEeqnarray}{rCl}
\sys{Y}_{s}&=&\overline{W}_{\sd{d}_k,h}\label{decode:Ys:a}\\
&&+ V_s+\mathop{\sum}\limits_{n\in[N]} \sys{p}_{k,n}\cdot\sys{W}_{n,h}\label{decode:Ys:b}\\
&&+\mathop{\sum}\limits_{\substack{(i,j)\in[F]\times[K]\\a_{i,j}=s,j\neq k}}\mathop{\sum}\limits_{n\in[N]}\sys{q}_{j,n}\cdot\sys{W}_{n,i}.\quad\label{decode:Ys:c}
\end{IEEEeqnarray}
\end{subequations}
For any $(i,j)\in[F]\times [K]$ such that $a_{i,j}=a_{h,k}=s$ and $j\neq k$, by the definition of PDA, we have $i\neq h$ and $a_{i,k}=*$. By~\eqref{eqn:cache:a}, this indicates that user $k$ has stored the  $i$-th packet of all files $\{\sys{W}_{n,i}:n\in[N]\}$. Moreover, by the fact that the user can get the coefficient vectors $\sd{q}_{[K]}$, user $k$ can compute the term in~\eqref{decode:Ys:c}. Furthermore, the signal in~\eqref{decode:Ys:b} is exactly the cached key $V_{a_{h,k}}\oplus T_{h,k}$ in~\eqref{eqn:cache:b} by~\eqref{eqn:privacy:key}. Thus, user $k$ can decode the packet $\overline{W}_{\sd{d}_k,h}$ by cancelling the terms in~\eqref{decode:Ys:b} and~\eqref{decode:Ys:c} in~\eqref{decode:Ys}.

\paragraph*{Security}
We prove the stronger condition in~\eqref{eqn:security:XW}, i.e.,
\begin{subequations}
\begin{IEEEeqnarray}{rCl}
&&I(\sd{d}_{[K]},W_{[N]};X)\\
&=&I(\sd{d}_{[K]},W_{[N]};\sd{q}_{[K]},Y_{[S]})\\
&=&I(\sd{d}_{[K]},W_{[N]};\sd{q}_{[K]})+I(\sd{d}_{[K]},W_{[N]};Y_{[S]}\,|\,\sd{q}_{[K]})\IEEEeqnarraynumspace\\
&=&0,\label{eq:exp5}
\end{IEEEeqnarray}
\end{subequations}
where~\eqref{eq:exp5} follows since (a) $\sd{q}_{[K]}=\sd{d}_{[K]} + \sd{p}_{[K]}$ is independent of $(\sd{d}_{[K]},W_{[N]})$ because $\sd{p}_{[K]}$ are independently and uniformly distributed over $\mathbb{F}_q^N$; and
(b) $Y_{[S]} = V_{[S]} + Y_{[S]}'$ is independent of $(\sd{d}_{[K]},W_{[N]},\sd{q}_{[K]})$ because $V_{[S]} $ are independently and uniformly distributed over $\mathbb{F}_q^{B/F}$, where
\begin{IEEEeqnarray}{c}
Y_s'\triangleq \mathop{\sum}\limits_{\substack{(i,j)\in[F]\times[K]\\a_{i,j}=s}} \ \mathop{\sum}\limits_{n\in[N]}\sys{q}_{j,n}\cdot\sys{W}_{n,i}, \quad\forall\, s\in[S].
\end{IEEEeqnarray}

\paragraph*{Privacy} 
\rv{For any $\mathcal{S}\subseteq[K]$ with $\mathcal{S}\neq \emptyset$, we have}
\begin{subequations}
\begin{IEEEeqnarray}{rCl}
&&\rv{I(\sd{d}_{[K]\backslash\mathcal{S}};X,\sd{d}_{\mathcal{S}},Z_{\mathcal{S}}\,|\,W_{[N]})}\\
&=&\rv{I(\sd{d}_{[K]\backslash\mathcal{S}};\sd{d}_{\mathcal{S}}\,|\,W_{[N]})+I(\sd{d}_{[K]\backslash\mathcal{S}};X\,|\,\sd{d}_{\mathcal{S}},W_{[N]})}\notag\\
&&\rv{\quad+I(\sd{d}_{{[K]\backslash\mathcal{S}}}; Z_\mathcal{S}\,|\,\sys{W}_{[N]}, \sys{X}, \sd{d}_\mathcal{S})}\\
  &=&I(\sd{d}_{{[K]\backslash\mathcal{S}}}; Z_\mathcal{S}\,|\,\sys{W}_{[N]}, \sys{X}, \sd{d}_\mathcal{S})\label{eqn:step:drop}\\
   &=& I(\sd{d}_{{[K]\backslash\mathcal{S}}}; Z_{\mathcal{S}} \,|\,\sys{W}_{[N]}, \sd{q}_{[K]}, \sd{d}_\mathcal{S},Y_{[S]}) \\
  &\leq& I(\sd{d}_{{[K]\backslash\mathcal{S}}}; Z_{\mathcal{S}},    V_{[S]} \,|\,\sys{W}_{[N]}, \sd{q}_{[K]}, \sd{d}_\mathcal{S},Y_{[S]})\\
   &=& I(\sd{d}_{{[K]\backslash\mathcal{S}}};  Z_\mathcal{S} ,V_{[S]}\,|\,\sys{W}_{[N]},\sd{q}_{[K]\backslash\mathcal{S}}, \sd{d}_\mathcal{S},\sd{p}_\mathcal{S},Y_{[S]})\label{eqn:step:add:explan}\\
   &=& I(\sd{d}_{{[K]\backslash\mathcal{S}}};  V_{[S]}\,|\,\sys{W}_{[N]}, \sd{q}_{[K]\backslash\mathcal{S}}, \sd{d}_\mathcal{S},\sd{p}_\mathcal{S},Y_{[S]})\IEEEeqnarraynumspace\label{eqn:drop:ZS}\\
   & =& 0,
\end{IEEEeqnarray}
\end{subequations}
where~\eqref{eqn:step:drop} follows from~\eqref{eqn:request} and~\eqref{eq:exp5};         ~\eqref{eqn:step:add:explan} follows since $\sd{q}_{\mathcal{S}}$ and $\sd{p}_{\mathcal{S}}$ determines each other given $\sd{d}_{\mathcal{S}}$ by~\eqref{eqn:bk};   and~\eqref{eqn:drop:ZS} follows since $Z_{\mathcal{S}}$ is determined by $\mathbf{p}_{\mathcal{S}},V_{[S]}$ and $W_{[N]}$ by construction.

\paragraph*{Performance}
By~\eqref{eqn:pda:packets}, each file is split into $F$ equal-size packets, each of length $\frac{B}{F}$, thus the subpacketization is $F$. For each user $k\in[K]$, by the cached content in~\eqref{eqn:cache}, for each $i\in[F]$ such that $a_{i,k}=*$, there are $N$ associated packets cached by the user, one from each file (see~\eqref{eqn:cache:a}). For each $i\in[F]$ such that $a_{i,k}\neq *$, there is one associated coded packet cached at the user (see~\eqref{eqn:cache:b}). Recall that, each column of a $(K,F,Z,S)$ PDA has $Z$ $``*"$s and $F-Z$ ordinary symbols, thus, the cache size at each user is
\begin{IEEEeqnarray}{c}
M_\mathbf{A}=\frac{1}{B}(Z\cdot N+F-Z)\frac{B}{F}=\frac{F+Z\cdot(N-1)}{F}.\IEEEeqnarraynumspace
\end{IEEEeqnarray}
By~\eqref{eqn:X}, the server sends $S$ coded packets $Y_{[S]}$, each of $\frac{B}{F}$ symbols, and the coefficient vectors  $\sd{q}_{[K]}$ can be sent in $KN$ symbols, thus the achieved load is
\begin{IEEEeqnarray}{c}
R_{\mathbf{A}}=\liminf_{B\rightarrow\infty}\frac{1}{B}\bigg(\frac{S\cdot B}{F}+KN\bigg)=\frac{S}{F}.
\end{IEEEeqnarray}

\begin{remark}[On randomness needed at the server]
%
One important assumption of all secure or private schemes known in the literature, including our novel SP-LFR scheme, is the availability of randomness at the server. In our problem formulation, this is represented by the random variable $\sys{P}$ from some unconstrained probability space $\mathcal{P}$ (see Section~\ref{sec:model}), which can be used by the server to privately generate whatever it is needed to guarantee privacy (see Section~\ref{sec:scheme}). In all known schemes, this boils down to the ability of the server to access \emph{perfectly random bits}, that is, i.i.d. uniformly distributed bits. This is a very strong assumption, as what one can practically have are just pseudo-random number generators \cite{VonNeuman1951}. For example, our $(K,F,Z,S)$ PDA-based SP-LFR requires $H(P) = (\frac{SB}{F}+NK)\log_2q$ perfectly random bits; the virtual users scheme requires $H(P) = K\log_2 N$ perfectly random bits; the scheme with random permutations in \cite{Kai2019Private} requires $H(P) = K\log_2\big({NK\choose t}!\big)$ perfectly random bits when $M=\frac{t}{K}$ for $t\in[0:NK]$;\footnote{The setup in \cite{Kai2019Private} considers a case where each user requires $L$ out of the $N$ files, in which case it requires $H(P)=N\log_2\big({U\choose t}!\big)$ perfect random bits when $M=\frac{tN}{U}$ for $t\in\big[0:U\big]$, where $U={N\choose L }K$.} the security key scheme in \cite{secure} requires $R_t B$  perfect random bits when the memory size is $M=M_t$ for $t\in[0:K]$, \rv{where $M_t$ and $R_t$ are given in~\eqref{eqn:MRt}}; and so on.
An interesting, and practically relevant question is thus: \emph{what is the minimum number of perfectly random bits needed to achieve privacy?}

As part of ongoing work we are investigating this question. The answer is clearly a tradeoff between load, memory and amount of randomness. To see this, consider the P-LFR setup. On the one hand, by prefetching arbitrary $MB$  symbols of the files to the cache of each user, the server can send all the remaining $(N-M)B$ symbols to the users without any coding, thus achieving the memory-load pair $(M,N-M)$ which does not require any randomness at the server. On the other hand, if we remove the security key in MAN-PDA-based SP-LFR scheme, we obtain a P-LFR  scheme achieving the memory-load pair $(M_t,R_t)$, given in Corollary~\ref{cor:MAN}, if the server has $H(P)=KN\log_2q$ perfectly random bits.
\end{remark}

\begin{remark}[On the Updated of Superposition Keys]\label{remark:update:keys} 
\rv{

The superposition keys stored in the caches can only be used once, i.e., similarly to a Shannon's one-time-pad which compromises privacy if re-used~\cite{ShannonOneTimePad}. The issue is that in our scheme the privacy key is hard-coded (in a multiplicative manner with the files) and then pre-fetched; thus the prefetched material needs to be updated from the server each time after users request a linear combination of files; this requires the prefetching of $1 - \frac{M}{N}$ 
files from the server each time a request needs to be satisfied. This implies that only a single demand can be satisfied between a placement phase and the subsequent one, which may make the system not practical\footnote{\rv{One important role of caching in practical system is to shift the communication load of the network from peak traffic times to off-peak traffic times, where the communication cost at off-peak times is quite cheap. That is the motivation of the assumption that the  placement phase (off-peak times) is accomplished at no cost, and the major objective is to reduce the communication load in the delivery phase (peak times)~\cite{Maddah-Ali2014fundamental}. This makes it possible to update the superposition key periodically at off-peak times at low cost in practical systems.}
}.

If multiple requests need to be satisfied in between two consecutive placements phases, a practical solution could be to use local randomness (in practice through a local pseudo-random generator) and allow vanishing security and privacy.
Users can update their cached content after every delivery by public randomness (known by the server and all users) and their local randomness (not known by the other users but known by the server) and what was delivered by the server. The process for the  SP-LFR scheme based on a $(K,F,Z,S)$ PDA $\sd{A}=[a_{i,j}]_{F\times K}$ is as follows. 
\begin{itemize}

\item
Initialization. The files are partitioned as in \eqref{eqn:pda:packets}.  
The server generates $K$  i.i.d. vectors $\sd{p}^0_k,k\in[K]$ uniformly in $\mathbb{F}_q^N$  and $S$ vectors $V_{s}^{0},s\in[S]$ uniformly in $\mathbb{F}_q^{B/F}$. The initial cache content of user $k$ is
\begin{subequations}
\begin{IEEEeqnarray}{rCl}
Z^{0}_k& = 
  &\big\{W_{n,i} : i\in[F],a_{i,k}=*,n\in[N]\big\}   \label{eq:new:cacheuncoded}   
\\&&\cup \big\{V_{a_{i,k}}^0 +\overline{W}_{\sd{p}^{0}_{k}, i} : i\in[F],a_{i,k}\neq *\big\}, \label{eq:new:cachecoded} 
\end{IEEEeqnarray}
\end{subequations}
where~\eqref{eq:new:cacheuncoded} is the uncoded cached content which will not change, and~\eqref{eq:new:cachecoded} is a linear combination of files that will be updated at end end of every delivery round.

\item
In delivery round $\ell=0,1,2,\ldots$ user $k\in[K]$ requests $\sd{d}^{\ell}_{k}$. 
The server will satisfy user $k$ with the LFR demands $\sd{q}^{\ell}_{k} =  \sd{p}^{\ell}_{k} + \sd{d}^{\ell}_{k}$. The server sends all the coefficients $\sd{q}_{[K]}^\ell$ and multicast messages $Y_s^{\ell}$ ($s\in[S]$) to the users, where $Y_s^{\ell}$ is created according to \eqref{eqn:pda:Ys}, with $V_s$ and $\sd{q}_k$ substituted by $V_s^{\ell}$ and $\sd{q}_k^\ell$ respectively. 
%
%

\item
At the end of the $\ell$-th delivery round, the cache content of user $k\in[K]$ is updated by using the retrieved linear combinations 
$W_{\sd{d}_k}$, a coefficient $c_k^{\ell}\in\mathbb{F}_q$ generated locally uniformly at random and a set of new public i.i.d.  vectors $\{V_s^{\rm{u},\ell}:s\in[S]\}$ uniformly distributed over $\mathbb{F}_q^{B/F}$. The term in~\eqref{eq:new:cachecoded} is replaced by 
\begin{IEEEeqnarray}{rCl}
&&\underbrace{ V_{a_{i,k}}^{\ell}+\overline{W}_{\sd{p}^{\ell}_{k}, i} }_\text{old superposition key} + 
\underbrace{V_{a_{i,k}}^{\rm{u},\ell}+c^{\ell}_{k} \cdot \overline{W}_{\sd{d}^{\ell}_{k}, i}}_\text{update}
\label{eq:new:cacheupdate}\\
&=&\underbrace{ V_{a_{i,k}}^{\ell}+V_{a_{i,k}}^{\rm{u},\ell} }_\text{$V_{a_{i,k}}^{\ell+1}$} + 
\underbrace{\overline{W}_{\sd{p}^{\ell}_{k}, i}+c^{\ell}_{k} \cdot \overline{W}_{\sd{d}^{\ell}_{k}, i}}_\text{$\overline{W}_{\sd{p}^{\ell+1}_k, i}$}.
\end{IEEEeqnarray}
That is, in the $(\ell+1)$-th delivery round, in the components of the superposition keys, the security keys are updated as
\begin{IEEEeqnarray}{c}
V_s^{\ell+1}=V_s^\ell+V_s^{\rm{u},\ell},\quad\forall\, s\in[S],
\end{IEEEeqnarray}
while the vectors $\sd{p}_k$ ($k\in[K]$)  generating the privacy keys are  shifted to a random distance along the direction $\sd{d}_k^{\ell}$, which is not known by the other users, i.e., 
\begin{IEEEeqnarray}{c}
  \sd{p}^{\ell+1}_{k} = \sd{p}^{\ell}_{k} + c^{\ell}_{k} \cdot \sd{d}^{\ell}_{k}, 
\quad\forall\,k\in[K].
  \label{eq:new:keyupdate}
\end{IEEEeqnarray}
\end{itemize}
In practical systems, the updated components of security keys $V_s^{\rm{u},\ell}$ can be generated using pseudo-random keys and delivered by the server. Thus, the  superposition keys can be updated with a cheap cost,  although the perfect security/privacy is  not achieved. 
}

%
%

\end{remark}

\section{Lower Bound \rv{for} PDA-based SP-LFR Schemes and Optimality of MAN-PDA (Proof of Theorem~\ref{thm:MAN:opt:PDA})}
In this section, we first present two known useful properties of PDAs, and then prove Theorem~\ref{thm:MAN:opt:PDA}.

\subsection{Preliminary Lemmas}

\begin{lemma}[Lemma 3 of~\cite{ChengTcomPDA}]\label{lemma:S:lowerbound} Given any $F\times K$ array $\mathbf{A}$, whose entries are composed of a specific symbol $``*"$ and some ordinary symbols, denoted by $1,2,\ldots,S$, if $\mathbf{A}$ satisfies the condition that two distinct entries $a_{i,j}=a_{i',j'}=s$ for some ordinary symbol $s\in[S]$ only if the conditions \textnormal{a)} and \textnormal{b)} in Definition~\ref{def:PDA} hold, then
\begin{IEEEeqnarray}{c}
S\geq\frac{nF}{KF+F-n},\label{eqn:S:bound}
\end{IEEEeqnarray}
where the number of ordinary entries is denoted $n$.
Moreover, the inequality in~\eqref{eqn:S:bound} holds with equality if and only if there are $\frac{n}{F}$ ordinary symbol entries in each row, and each symbol $s\in[S]$ occurs $\frac{n}{S}$ times.
\end{lemma}

\begin{lemma}[Lemma 2 of~\cite{Yan2017PDA}]\label{lemma:Yan} Given any positive integers $K,F,g$ such that $K\geq g\geq 2$, if an $F\times K$ array $\mathbf{A}$ whose entries are composed of a specific symbol $``*"$ and some ordinary symbols, denoted by $1,2,\ldots,S$,  if it satisfies the following conditions:
\begin{enumerate}
  \item each row has exactly $g-1$ $``*"$s;
  \item each ordinary symbol occurs exactly $g$ times; and
  \item two distinct entries $a_{i,j}=a_{i',j'}=s$ for some ordinary symbol $s\in[S]$ only if the conditions \textnormal{a)} and \textnormal{b)} in Definition~\ref{def:PDA} hold;
\end{enumerate}
then $F\geq{K\choose g-1}$.
\end{lemma}


\subsection{Proof of Theorem~\ref{thm:MAN:opt:PDA}}
In the $(K,F,Z,S)$ PDA $\mathbf{A}$,  the number of ordinary symbols is given by $n=K(F-Z)$, thus by Lemma \ref{lemma:S:lowerbound}, we have
\begin{IEEEeqnarray}{rCl}
\frac{S}{F}&\geq&
\frac{K(F-Z)}{F+KZ}\label{SFbound:b}
=\frac{K(1-Z/F)}{1+KZ/F}.
\end{IEEEeqnarray}
By Theorem~\ref{thm:PDA}, an achieved $(M,R)$ pair satisfies
\begin{IEEEeqnarray}{c}
R=\frac{S}{F},\quad \frac{M-1}{N-1}=\frac{Z}{F}.\label{eqn:RMN}
\end{IEEEeqnarray}
Then the lower bound in~\eqref{eqn:PDA:bound} is directly obtained by plugging~\eqref{eqn:RMN} into~\eqref{SFbound:b}. The fact $\{(M_t,R_t):t\in[0:K]\}$ are on the curve of the lower bound can be straightly verified by  plugging $M=M_t$ and $R=R_t$ into~\eqref{eqn:PDA:bound}.

Assume that $(M,R)=(M_t,R_t)$, it remains to prove that $F\geq {K\choose t}$. In fact, for $t=0$ and $t=K$, the conclusion is  trivial. Now consider the case $t\in[K-1]$, the fact that the inequality in~\eqref{eqn:PDA:bound} holds with equality indicates that  the inequality in~\eqref{SFbound:b} holds with equality. Moreover, plugging the expression $M_t=1+\frac{t(N-1)}{K}$ into~\eqref{eqn:RMN}, we obtain
$
\frac{KZ}{F}=t
$,
and by~\eqref{SFbound:b} and the fact the inequality in~\eqref{SFbound:b} holds with equality,
$
\frac{K(F-Z)}{S}=t+1
$.
By Lemma~\ref{lemma:S:lowerbound}, there are $K-\frac{n}{F}=\frac{KZ}{F}=t$ $``*"$s in each row, and each symbol occurs $\frac{n}{S}=t+1$ times in the PDA $\mathbf{A}$. Therefore, the conclusion $F\geq{K\choose t}$ directly follows from Lemma~\ref{lemma:Yan}.

\section{Gap Results \rv{for} MAN-PDA-based SP-LFR Schemes (Proof of Theorem~\ref{thm:gap}) }\label{sec:gap}
In this section, we first present four useful lemmas, then we prove Theorem~\ref{thm:gap} by using these lemmas.  The proof of Lemma~\ref{lemma:converse},~\ref{lemma:MN:bound} and~\ref{lemma:bound:RMr} are deferred to Appendix~\ref{app:lemma:converse},~\ref{app:MN:bound} and~\ref{app:bound:RMr}, respectively. Lemma~\ref{lemma:fact2} directly follows from the result of~\cite{QYu2018Factor2}.

\subsection{Preliminary Lemmas}
%

\begin{lemma}[Cut-set bound]\label{lemma:converse} The optimal SP-LFR load-momery tradeoff $R^*(M)$ \rv{satisfies}
\begin{IEEEeqnarray}{c}
R^*(M)\geq \max_{u\in[\min\{\lfloor\frac{N}{2}\rfloor,K\}]}\frac{uN-u^2M}{N-1},\quad\forall\, M\in[1,N].\IEEEeqnarraynumspace
\end{IEEEeqnarray}
\end{lemma}

The cut-set bound here is derived similarly to~\cite{secure}, but with some different steps in the proof (see Remark \ref{remark:cutset}).  The advantage of the cut-set bound in Lemma \ref{lemma:converse} is that it does not contain rounding operations such as $\lfloor \frac{N}{s}\rfloor$. The technique of removing rounding here is different from but simpler than the approach based on Han's inequality in~\cite{KaiCacheCommun2019}. 

\begin{lemma}\label{lemma:MN:bound}
The load of the MAN-PDA-based SP-LFR scheme  satisfies
\begin{IEEEeqnarray}{c}
R_{\textnormal{MAN}}(M)\leq\frac{N-M}{M-1},\quad\forall\, M\in(1,N].\label{ineq:RM}
\end{IEEEeqnarray}
\end{lemma}

Let $r_{\textnormal{MAN}}(M)$ be the lower convex envelope of the following points
\begin{IEEEeqnarray}{c}
M_t'=\frac{tN}{K},\quad R_t'=\frac{K-t}{t+1},\quad t\in[0:K].\label{eqn:RMtprime}
\end{IEEEeqnarray}
Then, for $N\geq K$, $r_{\textnormal{MAN}}(M)$ is the optimal worst-case load-memory tradeoff under the constraint of uncoded placement in the FR setup~\cite{Kai2020Index,Yu2019ExactTradeoff}.

 \begin{lemma}\label{lemma:bound:RMr} If $N\geq K\geq 2$, the loads $R_{\textnormal{MAN}}(M)$ and $r_{\textnormal{MAN}}(M)$ satisfy
\begin{IEEEeqnarray}{rCl}
\frac{R_{\textnormal{MAN}}(M)}{r_{\textnormal{MAN}}(M)}&\leq&\left\{\begin{array}{cc}
                                                     2 , &\textnormal{if}~N\geq K+2  \\
                                                     2.5,  &\textnormal{if}~N= K+1 \\
                                                     3,  &\textnormal{if}~N=K\geq 3
                                                           \end{array}
\right.,~\forall  M\in[1,N].\IEEEeqnarraynumspace\label{eqn:Rr:bound}
\end{IEEEeqnarray}
\end{lemma}

\begin{lemma}[Theorem 1 in~\cite{QYu2018Factor2}]\label{lemma:fact2}
Let $r_{\textnormal{FR}}^*(M)$ be the optimal worst-case load-memory tradeoff in the FR setup.
For $N\geq K$, $r_{\textnormal{MAN}}(M)$ satisfies
\begin{IEEEeqnarray}{c}
\frac{r_{\textnormal{MAN}}(M)}{r_{\textnormal{FR}}^*(M)}\leq 2.00884,\quad\forall\, M\in[0,N].\label{eqn:r:bound}
\end{IEEEeqnarray}\end{lemma}

\subsection{Proof of Theorem~\ref{thm:gap}}
We separately prove the case $N\geq K$ and the case $ N < K, \ M\geq 2$.

\subsubsection{Case $N\geq K$}
If $K=1$, we have $R^*(M)\geq 1-\frac{M-1}{N-1}$ by Lemma \ref{lemma:converse} and  $R_{\textnormal{MAN}}(M)=1-\frac{M-1}{N-1}$ by Corollary \ref{cor:MAN}, thus $R_{\textnormal{MAN}}(M)=R^*(M)$.  
If $N=K=2$, then $R_{\textnormal{MAN}}(M)$ is obtained by sequentially connecting the points $(M_0,R_0)=(1,2), (M_1,R_1)=(\frac{3}{2},\frac{1}{2}), (M_2,R_2)=(2,0)$, i.e.,
\begin{IEEEeqnarray}{c}
R_{\textnormal{MAN}}(M)=\left\{\begin{array}{ll}
                            5-3M, &\textnormal{if}~1\leq M\leq\frac{3}{2}  \\
                            2-M, &\textnormal{if}~\frac{3}{2}\leq M\leq 1
                          \end{array}
\right.
\end{IEEEeqnarray}
and by Lemma~\ref{lemma:converse},
\begin{IEEEeqnarray}{c}
R^*(M)\geq 2-M,\quad \forall\, M\in[1,2].
\end{IEEEeqnarray}
Therefore,
\begin{IEEEeqnarray}{c}
\frac{R_{\textnormal{MAN}}(M)}{R^*(M)}\leq \max_{x\in [1,\frac{3}{2}]}\Big\{\frac{5-3x}{2-x},1\Big\}=2,\quad \forall M\in[1,2].\IEEEeqnarraynumspace\label{case1}
\end{IEEEeqnarray}
If $N>2$ or $K>2$, since the optimal load in FR setup does not exceed the optimal load in SP-LFR setup,  it holds  that $r_{\textnormal{FR}}^*(M)\leq R^*(M)$.
Hence,  by the bounds in~\eqref{eqn:Rr:bound} and~\eqref{eqn:r:bound}, for any $M\in[1,N)$ we have
\begin{IEEEeqnarray}{rCl}
\frac{R_{\textnormal{MAN}}(M)}{R^*(M)}&=&\frac{R_{\textnormal{MAN}}(M)}{r_{\textnormal{MAN}}(M)}\cdot \frac{r_{\textnormal{MAN}}(M)}{r_{\textnormal{RF}}^*(M)}\cdot\frac{r_{\textnormal{RF}}^*(M)}{R^*(M)}\\
&\leq&\left\{\begin{array}{ll}
               4.01768, & \textnormal{if}~N\geq K+2 \\
               5.0221, & \textnormal{if}~N=K+1 \\
               6.02652, &\textnormal{if}~N=K\geq 3
             \end{array}
\right..\label{case2}
\end{IEEEeqnarray}
By combining~\eqref{case1} and~\eqref{case2}, the case $N\geq K$ is proved.

\subsubsection{$N < K$ and $2\leq M < N$}  For each $u\in\left[ \lfloor\frac{N}{2}\rfloor \right]$, define
\begin{IEEEeqnarray}{c}
L_u(M) \triangleq \frac{uN-u^2M}{N-1},\quad M\in[0,N].
\end{IEEEeqnarray}
Notice that $\lfloor \frac{N}{2}\rfloor <K$, so for each $u\in\left[ \lfloor\frac{N}{2}\rfloor \right]$, by Lemma~\ref{lemma:converse},
\begin{IEEEeqnarray}{c}
R^*(M)\geq L_u(M),\quad\forall\, M\in[1,N].\label{ineq:RLs}
\end{IEEEeqnarray}

For any $M\in\big[\frac{N}{2\lfloor N/2\rfloor+1},N\big]$, define
\begin{IEEEeqnarray}{c}
f(M)\triangleq\frac{1}{4}\cdot\frac{N}{N-1}\Big(\frac{N}{M}-\frac{M}{N}\Big).
\end{IEEEeqnarray}
Notice that by the fact $\lfloor \frac{N}{2}\rfloor\geq \frac{N-1}{2}$, $\frac{N}{2\lfloor N/2\rfloor+1}\leq 1$. Thus, the interval $\big[\frac{N}{2\lfloor N/2\rfloor+1},N\big]$  encloses $[1,N]$ as its sub-interval. We claim that $f(M)$ lower bounds $R^*(M)$ on $[1,N]$, i.e.,
\begin{IEEEeqnarray}{c}
R^*(M)\geq f(M),\quad\forall\, M\in[1,N].\label{eqn:Rf:bound}
\end{IEEEeqnarray}
In fact, consider the interval $\big[\frac{N}{2\lfloor N/2\rfloor+1},N\big]$, which can be split into the $\lfloor \frac{N}{2}\rfloor$ intervals as
  \begin{IEEEeqnarray}{c}
  \Big[\frac{N}{2\lfloor N/2\rfloor+1},N\Big]=\bigcup_{u=1}^{\lfloor \frac{N}{2}\rfloor}\Big[\frac{N}{2u+1},\frac{N}{2u-1}\Big].\label{eqn:interval:partition}
  \end{IEEEeqnarray}
For any $M\in[1,N]$, there exists  $u\in\left[ \lfloor\frac{N}{2}\rfloor \right]$ such that $M\in[\frac{N}{2u+1},\frac{N}{2u-1}]$. It is easy to verify the following two equalities:
\begin{subequations}\label{eqn:convex}
\begin{IEEEeqnarray}{rCl}
L_u\left(\frac{N}{2u+1}\right)&=&f\left(\frac{N}{2u+1}\right)=\frac{N}{N-1}\cdot\frac{u(u+1)}{2u+1},\IEEEeqnarraynumspace\\
L_u\left(\frac{N}{2u-1}\right)&=&f\left(\frac{N}{2u-1}\right)=\frac{N}{N-1}\cdot\frac{u(u-1)}{2u-1}.\IEEEeqnarraynumspace
\end{IEEEeqnarray}
\end{subequations}
Since $f(x)$ is convex on the interval $[\frac{N}{2u+1},\frac{N}{2u-1}]$, by~\eqref{eqn:convex},
\begin{IEEEeqnarray}{c}
L_u(x)\geq f(x),\quad \forall\, x\in\left[\frac{N}{2u+1},\frac{N}{2u-1}\right].\label{ineq:Lsf}
\end{IEEEeqnarray}
Hence, by~\eqref{ineq:RLs},~\eqref{eqn:interval:partition} and~\eqref{ineq:Lsf}, we conclude that~\eqref{eqn:Rf:bound} holds.
Moreover, for any $M\in[2,N)$, by~\eqref{ineq:RM} and~\eqref{eqn:Rf:bound},
\begin{subequations}
\begin{IEEEeqnarray}{rCl}
&\frac{R_{\textnormal{MAN}}(M)}{R^*(M)}
&\leq
   \frac{\frac{N-M}{M-1}}{\frac{1}{4}\cdot\frac{N}{N-1}\cdot\Big(\frac{N}{M}-\frac{M}{N}\Big)}\\
&=&4\cdot\frac{N-1}{N}\cdot\frac{N-M}{M-1}\cdot\frac{MN}{(N+M)(N-M)}\\
&=&4(N-1)\cdot\Big(\frac{1}{M+N}+\frac{1}{(M-1)(M+N)}\Big)\IEEEeqnarraynumspace\\
&\leq&8\cdot\frac{N-1}{N+2}\\
& < &8.
\end{IEEEeqnarray}
\end{subequations}
\rv{This concludes the proof.}

\section{Numerical Results}\label{sec:numerical}

\begin{figure}[htbp] \centering
\subfigure[$N=30,K=10$] {
 \label{fig2:a}
\includegraphics[width=0.85\columnwidth]{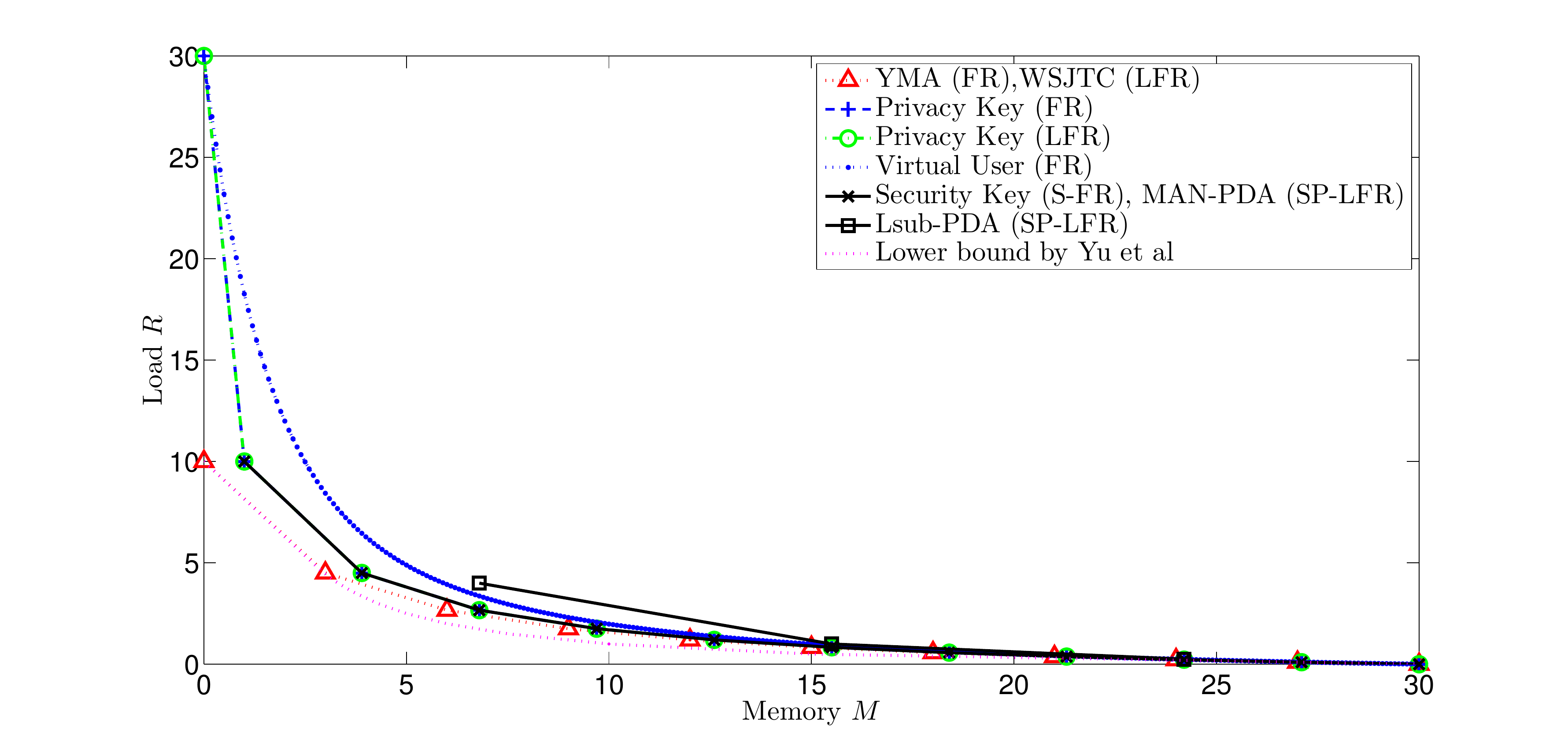}
}
\subfigure[$N=K=20$] {
\label{fig2:b}
\includegraphics[width=0.85\columnwidth]{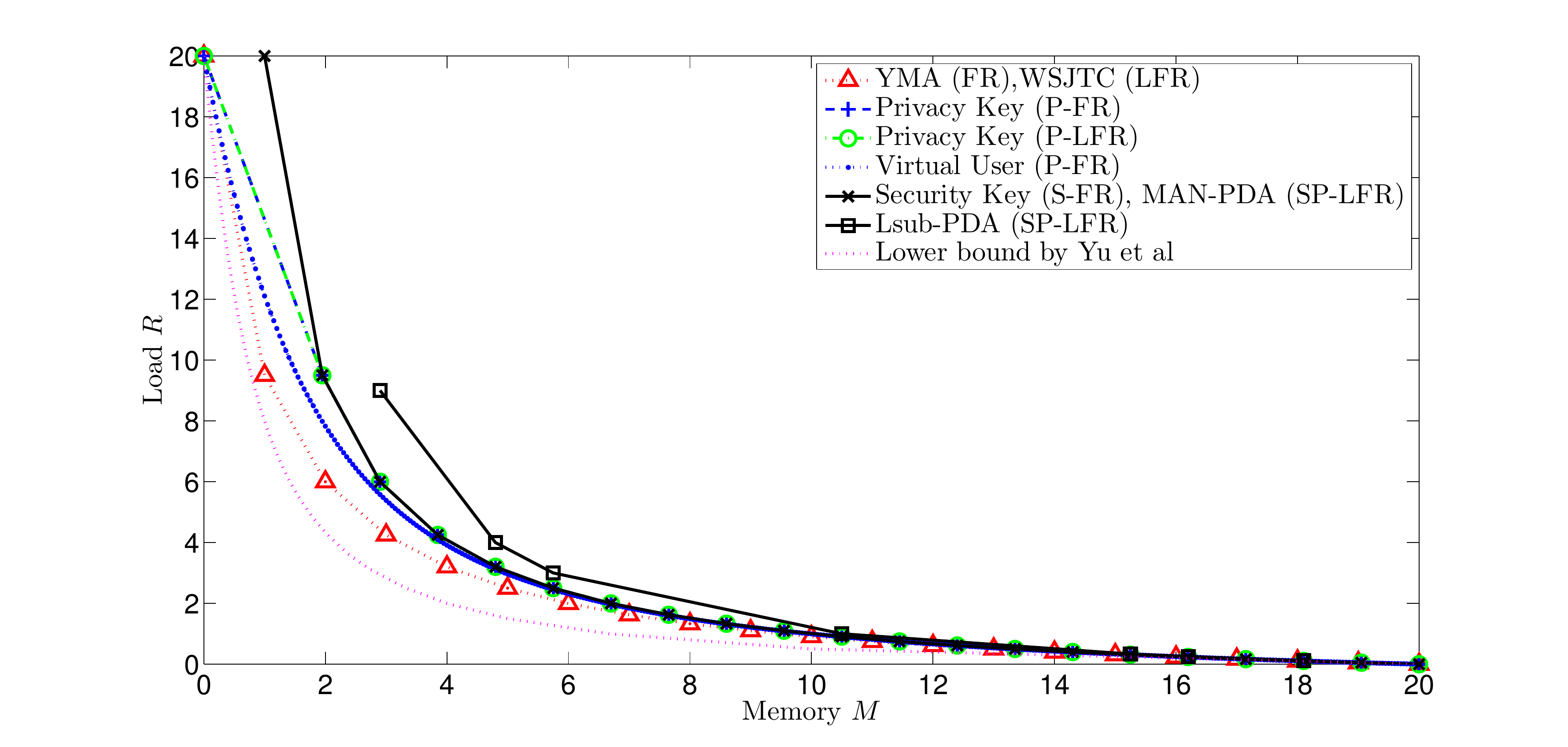}
}
\subfigure[$N=10,K=30$] {
\label{fig2:c}
\includegraphics[width=0.85\columnwidth]{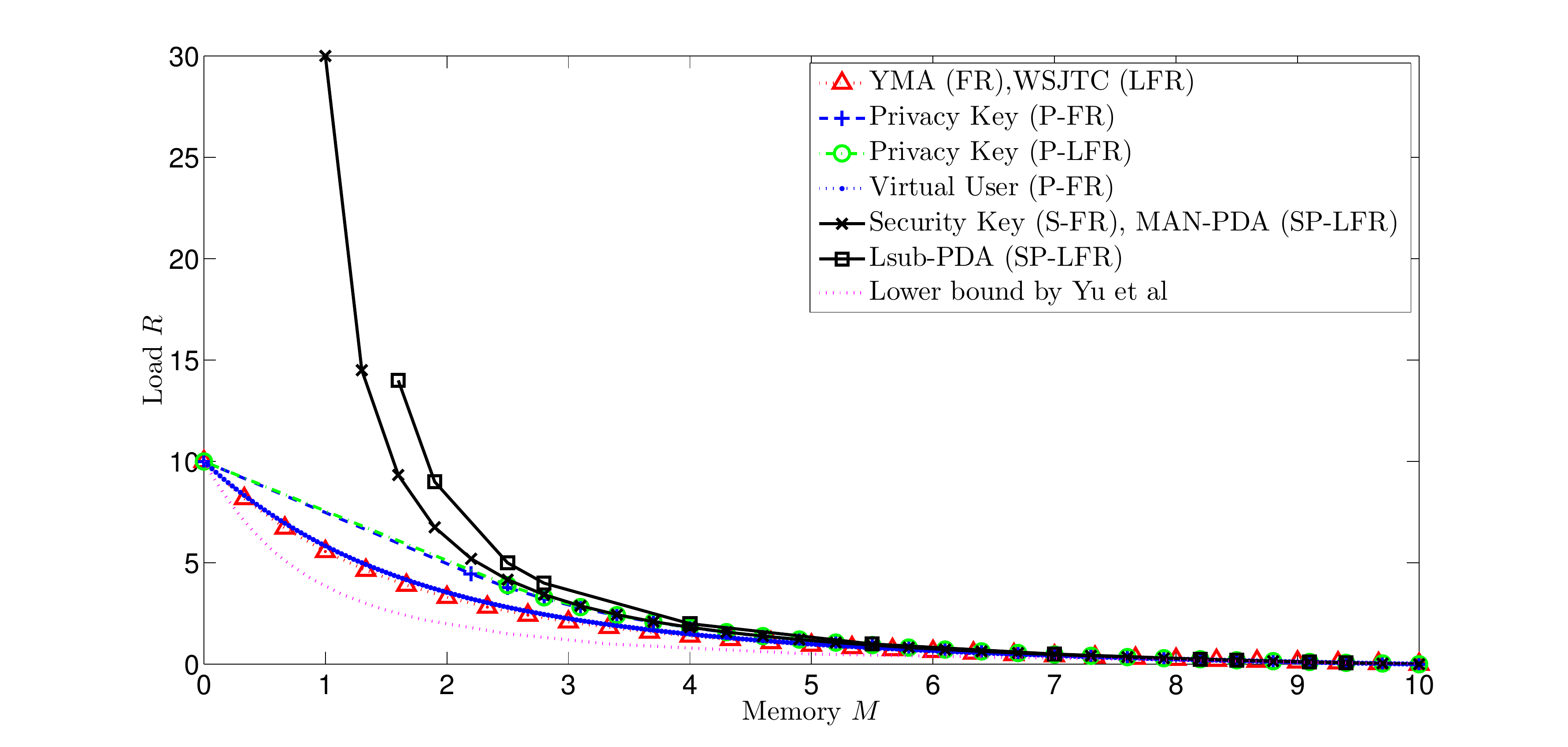}
}
\caption{Load-memory tradeoff for secure/non-secure and private/non-private systems such that (a) $N>K$; (b) $N=K$; (c) $ N<K$. }
\label{fig:2}
\end{figure}

In this section, we compare numerically the performance of the schemes in Corollary~\ref{cor:MAN} and~\ref{corollary:PDA}, with  the schemes listed in Table~\ref{table:compare}. For simplicity, we will refer to the PDA based schemes in Corollary~\ref{cor:MAN} and~\ref{corollary:PDA} as MAN-PDA and Lsub-PDA SP-LFR schemes, respectively. In Fig.~\ref{fig:2}, we plot the memory-load tradeoff of these schemes in three regimes: $N>K$, $N=K$ and $N < K$, where we choose parameters $(N,K)=(30,10),(20,20)$ and $(10,30)$, respectively. For reference, we also plot the converse bound in~\cite{QYu2018Factor2}, which works for all the above schemes.

In Fig.~\ref{fig:2}, comparing the MAN-PDA-based SP-LFR scheme with the other schemes, we make the following observations.
\begin{enumerate}

  \item In all cases, the MAN-PDA-based SP-LFR scheme achieves the same performance as the security key scheme in the S-FR setup. In fact, the superposition of the security keys and privacy keys, as well as the extension form \rv{single} file retrieval to linear function retrieval, increases neither the memory nor the load.

  \item For the case $N>K$ (Fig.~\ref{fig2:a}), the MAN-PDA-based SP-LFR scheme achieves the same performance as the privacy key schemes (in both P-FR and P-LFR setups) on the whole interval $[1,N]$. In fact, for the chosen parameter, the load-memory tradeoff of the privacy key schemes on the interval $M\in[1,N]$ are given by the lower convex envelope of the points $\{(M_t,R_t):t\in[0:K]\}$ \rv{in~\eqref{eqn:MRt}}, which is exactly $R_{\textnormal{MAN}}(M)$. 
  Again, the superposition of the privacy keys with the security keys increases neither the memory size nor the load.
      It was numerically verified in~\cite{LFR:DPCU} that when $N>2K+1$ and $0<M<N-1-\frac{1}{K}$, the privacy key scheme  outperforms the virtual users scheme in the P-FR setup. Thus, in the regime $N>2K+1,1\leq M\leq N-1-\frac{1}{K}$, the MAN-PDA-based SP-LFR scheme also outperforms the virtual users scheme.

  \item For the case $N\leq K$ (Fig.~\ref{fig2:b} and~\ref{fig2:c}), the MAN-PDA-based SP-LFR scheme achieves the same performance as the privacy key schemes when $M$ is lager than $1+\frac{(K-N+1)(N-1)}{K}$. For small $M$, it is inferior to the privacy key scheme because in the P-FR setup:
      \begin{enumerate}
        \item the trivial point $(M,R)=(0,N)$ can be achieved, and thus memory-sharing the other points with this point increases the performance; and
        \item 
        some redundant signals can be removed, similar to the cases in~\cite{QYu2018Factor2} and~\cite{Kai2020LinearFunction}.
      \end{enumerate}
      For SP-LFR systems, the above two points do not hold due to the use of security keys in the \rv{delivery} signals.
       Notice that, at the corner points, the additional load in the MAN-PDA-based SP-LFR scheme  due to the redundant signals compared to the privacy key scheme in P-FR setup is ${{K-\min\{N-1,K\}\choose t+1}}/{{K\choose t}}$, where $t=K\frac{M-1}{N-1}\in[0:K-N]$. This indicates that when  $N<K$ and $M$ is close to $1$, the additional load is significant (e.g., at $M=1$, the additional load is $K-N+1$), which leads to the observation that the load of MAN-PDA-based SP-LFR scheme diverges from that of the YMA/WSJTC/Privacy Key/Virtual users scheme in the FR/LFR setup. 


\end{enumerate}

We observe that the Lsub-PDA-based SP-LFR scheme in Corollary~\ref{corollary:PDA} is worse than, but close to, the load of the MAN-PDA-based SP-LFR scheme in all regimes. The advantage of the Lsub-PDA scheme is its low subpacketization. In fact, by Corollary~\ref{cor:MAN} and~\ref{corollary:PDA}, for integer $t\in[2:K-1]$ such that $t\,|\,K$, the MAN-PDA and Lsub-PDA-based SP-LFR schemes achieve the same memory size. Let us denote their loads and subpacketizations by $R_{\textnormal{MAN}},R_{\textnormal{Lsub}}$ and $B_{\textnormal{MAN}},B_{\textnormal{Lsub}}$ respectively.
Then,
\begin{IEEEeqnarray}{rCl}
R_{\textnormal{MAN}}&=&\frac{t}{t+1}R_{\textnormal{Lsub}},\label{load:cor12}\\
B_\textnormal{MAN}&=&\frac{K!}{t!(K-t)!}\\
&\geq&\frac{\sqrt{K}}{e^{\frac{1}{6}}\sqrt{2\pi t(K-t)}}\Big(\frac{K}{t}\Big)^t\Big(\frac{K}{K-t}\Big)^{K-t}\label{eqn:stirling}\\
&=&\frac{1}{e^{\frac{1}{6}}\sqrt{2\pi(K-t)}}\Big(\frac{K}{t}\Big)^{\frac{3}{2}}\Big(\frac{K}{A}\Big)^{K\cdot \frac{A}{K}} B_{\textnormal{Lsub}},\IEEEeqnarraynumspace\label{sim:stirling}
\end{IEEEeqnarray}
where $A\triangleq \max\{t,K-t\}$ and in~\eqref{eqn:stirling}, we used the Stirling's approximation $\sqrt{2\pi}n^{n+\frac{1}{2}}e^{-n}\leq n!\leq e^{\frac{1}{12}}\sqrt{2\pi}n^{n+\frac{1}{2}}e^{-n}$. From~\eqref{load:cor12},  the MAN-PDA- and Lsub-PDA-based SP-LFR schemes have similar loads. While from~\eqref{sim:stirling}, we see that if $t$ and $K$ increase proportionally\footnote{In this case, each user keeps the same memory size $1+\frac{t(N-1)}{K}$, and the number of users $K$ increases.},  the Lsub-PDA-based SP-LFR scheme  saves a factor that increases exponentially with $K$. It is worth pointing out that, the MAN-PDA-based SP-LFR scheme  keeps the same subpacketization as the privacy key scheme in Table~\ref{table:compare}, and thus also has great superiority over the virtual user scheme, as illustrated in ~\cite{LFR:DPCU}.

\section{Conclusions}\label{sec:conclusion}
In this paper, we investigated the cache-aided content Secure and demand Private Linear Function Retrieval (SP-LFR) problem, where the users are interested in decoding  linear combinations of the files, \rv{while the library content must be protected against a wiretapper observing the 
delivery signal, and the user demands must be protected against  the wiretapper and any subset of colluding users.}   We proposed to use a superposition of security keys and privacy keys to guarantee both content security and demand privacy. Moreover,  this idea was incorporated into the Placement Delivery Array (PDA) framework to obtain SP-LFR schemes from existing PDA results. In particular, among all PDA based SP-LFR schemes, the memory-rate pairs achieved by the PDAs that \rv{describe} Maddah-Ali and Niesen's coded caching scheme (MAN-PDA) are Pareto-optimal,  and they have the lowest subpacketization to achieve those points. Such strong optimality results on PDAs were not known in the coded caching literature, to the best of our knowledge.  In addition, the tradeoff was also showed to be optimal to within a constant multiplicative gap except for the regime $N<K,1\leq M\leq 2$. Remarkably, the MAN-PDA-based SP-LFR scheme does not increase the load compared to the best known S-FR schemes in all regimes, or the best known P-FR scheme in the regime \rv{where optimality guarantee can be proved}.

\begin{appendix}

\subsection{Proof of Lemma~\ref{lemma:converse}}\label{app:lemma:converse}

Consider the case where each user demand a file, and denote $D_k$ the index of the file demanded by user $k$ for all $k\in[K]$.
Denote the signal under the demands $D_1=d_1,$ $\ldots,$ $D_K=d_K$ by $X_{(d_1,\ldots,d_K)}$. For any $u\in \big[\min\{\lfloor \frac{N}{2}\rfloor,K\}\big]$, consider the first $u$ caches $Z_1,\ldots,Z_u$. For a demand vector
 $(D_1,\ldots,D_K)=(1,2,...,u,1,1,...,1)$, by using the signal $X_1\triangleq X_{(1,2,.\ldots,u,1,1,\ldots,1)}$ and the caches $Z_1,\ldots,Z_u$, the files $W_1,\ldots,W_u$ can be recovered. More generally, by using
 $X_{\ell}\triangleq X_{((\ell-1)u+1,\ldots,\ell\cdot u,1,1,\ldots,1)}$,
  the files $W_{(\ell-1)u+1},$ $W_{(\ell-1)u+2},$ $\ldots,$ $W_{\ell\cdot u}$ can be recovered for any $\ell\in\big[\lceil \frac{N}{u}\rceil-1\big]$.
   We also define
   $
   X_{\lceil\frac{N}{u}\rceil}\triangleq X_{((\lceil \frac{N}{u}\rceil-1)u+1,\ldots,N,1,1,\ldots,1)}$,
    then the files $W_{(\lceil \frac{N}{u}\rceil-1)u+1},\ldots,W_N$ can be recovered from $X_{\lceil\frac{N}{u}\rceil}$ and $Z_1,\ldots,Z_u$.
Therefore, since $W_1,\ldots,W_N$ are uniformly distributed over $\mathbb{F}_q^B$, then
\begin{IEEEeqnarray}{rCl}
NB&=&H(W_{[N]})\label{eqn:step1}\\
&=&I(W_{[N]};X_{[\lceil\frac{N}{s}\rceil]},Z_{[u]})+H(W_{[N]}\,|\,X_{[\lceil\frac{N}{u}\rceil]},Z_{[u]})\IEEEeqnarraynumspace\\
&=&I(W_{[N]};X_{[\lceil\frac{N}{u}\rceil]},Z_{[u]})\label{converse:exp1}\\
&=&I(W_{[N]};X_{\lceil\frac{N}{u}\rceil})+I(W_{[N]};X_{[\lceil\frac{N}{u}\rceil-1]},Z_{[u]}\,|\,X_{\lceil\frac{N}{u}\rceil})\IEEEeqnarraynumspace\\
&=&I(W_{[N]};X_{[\lceil\frac{N}{u}\rceil-1]},Z_{[u]}\,|\,X_{\lceil\frac{N}{u}\rceil})\label{eqn:step:98}\\
&\leq& \sum_{\ell=1}^{\lceil\frac{N}{u}\rceil-1}H(X_{\ell})+\sum_{j=1}^uH(Z_{u})\\
&\leq& \left(\left\lceil \frac{N}{u}\right\rceil-1\right) R^*(M)B+u MB\\
&\rv{\leq}&\frac{N-1}{u}R^*(M)B+uMB,\label{converse:exp2}
\end{IEEEeqnarray}
where~\eqref{converse:exp1} holds because  the files $W_{[N]}$ can be recovered from $X_{[\lceil\frac{N}{u}\rceil]},Z_{[u]}$;~\eqref{eqn:step:98} from the security condition~\eqref{eqn:security}; and~\eqref{converse:exp2} follows from the fact $\lceil\frac{N}{u}\rceil\leq \frac{N}{u}+\frac{u-1}{u}$. Therefore,
$
R^*(M)\geq \frac{uN-u^2M}{N-1}
$,
which works for all $u\in\big[\min\{\lceil\frac{N}{u}\rceil,K\}\big]$. This proves Lemma~\ref{lemma:converse}.

\begin{remark}[New ingredients in the cut-set bound]\label{remark:cutset}
The proof of this lemma is in fact a cut-set type bound, but it differs from~\cite{secure} in two steps.
(I) We use $\lceil \frac{N}{u}\rceil$ signals $X_1,\ldots,X_{\lceil\frac{N}{u}\rceil}$ to decode all the $N$ files, so that the left side in~\eqref{eqn:step1} is $NB$, while in~\cite{secure}, it used $\lfloor\frac{N}{u}\rfloor$ signals $X_1,\ldots,X_{\lfloor \frac{N}{u}\rfloor}$ to decode $\lfloor\frac{N}{u}\rfloor u$ files. This technique was also used in~\cite{Hierarchical}.
(II) In~\eqref{converse:exp2}, the inequality $\lceil\frac{N}{u}\rceil\leq \frac{N}{u}+\frac{u-1}{u}$ is used  so that the final lower bound does not contain the rounded number $\lceil \frac{N}{u}\rceil$.
 \end{remark}

\subsection{Proof of Lemma~\ref{lemma:MN:bound}}\label{app:MN:bound}
 Notice that the interval $(1,N]$ can be partitioned into $K$ disjoint intervals $\{(M_{t-1},M_t]:t\in[K]\}$.
Since the points $(M_0,R_0),(M_1,R_1),\ldots,(M_K,R_K)$ are on the convex curve
\begin{IEEEeqnarray}{c}
 \left(M,\frac{K(N-M)}{N-1+K(M-1)}\right), \quad M\in[0,N]
 \end{IEEEeqnarray}
 the lower convex envelope are formed by sequentially connecting the points  $(M_0,R_0),\ldots,(M_K,R_K)$.  For any $M\in (M_{t-1},M_t] $, there exists a unique $\theta\in[0,1)$ such that
\begin{IEEEeqnarray}{rCl}
M&=&\theta M_{t-1}+(1-\theta)M_t\\
&=&1+\frac{(t-\theta)(N-1)}{K},\IEEEeqnarraynumspace\label{eqn:M:th}\\
 R_{\textnormal{MAN}}(M)&=&\theta R_{t-1}+(1-\theta) R_t\\
 &=&\frac{(K-t)t+(K+1)\theta}{t(t+1)},\IEEEeqnarraynumspace\label{eqn:R:th}
 \end{IEEEeqnarray}
where~\eqref{eqn:M:th} and~\eqref{eqn:R:th} follow from~\eqref{eqn:MRt}.
 Therefore,
 \begin{IEEEeqnarray}{rCl}
 &&R_{\textnormal{MAN}}(M)-\frac{N-M}{M-1}\\
 &=&\frac{(K-t)t+(K+1)\theta}{t(t+1)}-\frac{K-t+\theta}{t-\theta}\label{eqn:step:96}\\
 &=&\frac{-(K-t)t-(K+1)\theta^2}{t(t+1)(t-\theta)} \\
 &\leq& 0,
 \end{IEEEeqnarray}
where~\eqref{eqn:step:96} follows from~\eqref{eqn:M:th},\eqref{eqn:R:th}.  
This proves~\eqref{ineq:RM}.

\subsection{Proof of Lemma~\ref{lemma:bound:RMr}}\label{app:bound:RMr}
Notice that since the points $(M_0',R_0'),$ $(M_1',R_1'),$ $\ldots,$ $(M_K',R_K')$ are on the convex curve
\begin{IEEEeqnarray}{c}
\Big(M,\frac{K(N-M)}{N+KM}\Big), \quad M\in[0,N],
\end{IEEEeqnarray}
the function $r_{\textnormal{RF}}(M)$ is given by sequentially connecting the points $(M_0',R_0'),\ldots,(M_K',R_K')$. Moreover, $R_{\textnormal{MAN}}(M)$ is convex on $[1,N]$, so it is sufficient to prove~\eqref{eqn:Rr:bound} for the points $M\in\{1\}\cup\big\{\frac{tN}{K}:t\in[K]\big\}$.

         If $M=1$, let $\theta=1-\frac{K}{N}$, then $M=1=\theta\cdot 0+(1-\theta)\frac{N}{K}$, thus
            \begin{IEEEeqnarray}{rCl}
            r_{\textnormal{MAN}}(1)&=&\theta\cdot r_{\textnormal{MAN}}(0)+(1-\theta)\cdot r_{\textnormal{MAN}}\Big(\frac{N}{K}\Big)\\
            &=&\Big(1-\frac{K}{N}\Big)\cdot K+\frac{K}{N}\cdot\frac{K-1}{2}\\
            &=&\frac{(2N-K-1)K}{2N},\label{eqn:rc}
            \end{IEEEeqnarray}
       Thus,
            \begin{IEEEeqnarray}{rCl}
            \frac{R_{\textnormal{MAN}}(1)}{r_{\textnormal{MAN}}(1)}&=&\frac{K}{\frac{(2N-K-1)K}{2N}}
           \\
           & =&\frac{2}{2-\frac{K+1}{N}}\\
            &\leq&\left\{\begin{array}{ll}
                          2, &\textnormal{if}~N\geq K+1  \\
                          3, &\textnormal{if}~N=K\geq 3
                         \end{array}
            \right.,
            \end{IEEEeqnarray}
             which satisfies~\eqref{eqn:Rr:bound}.

              If $M=\frac{tN}{K}$, where  $t\in[K]$, let $\theta_t=\frac{K-t}{N-1}\in[0,1]$, then
              \begin{IEEEeqnarray}{c}
             M=\theta_t M_{t-1}+(1-\theta_t)M_t.
             \end{IEEEeqnarray}
             Thus,
              \begin{IEEEeqnarray}{rCl}
              \frac{R_{\textnormal{MAN}}(M)}{r_{\textnormal{MAN}}(M)}&=&\frac{\theta_t R_{t-1}+(1-\theta_t) R_t}{R_t'}\\
              &=&1+\frac{K+1}{(N-1)t}\label{eqn:step:107}\\
              &\leq&1+\frac{K+1}{N-1}\\
              &\leq&\left\{\begin{array}{ll}
                            2,  &\textnormal{if}~N\geq K+2  \\
                            2.5, & \textnormal{if}~N=K+1 \\
                            3, & \textnormal{if}~N=K\geq 3
                           \end{array}
              \right.,\label{eqn:step:110}
              \end{IEEEeqnarray}
             where~\eqref{eqn:step:107} follows from~\eqref{eqn:MRt},\eqref{eqn:RMtprime}; and in~\eqref{eqn:step:110}, we used the fact $K\geq 2$. This proves~\eqref{eqn:Rr:bound}.

\end{appendix}

\end{document}